# Completion of Lunar Magma Ocean Solidification at 4.43 Ga

Nicolas Dauphas[1*], Zhe J. Zhang[1], Xi Chen[1], Mélanie Barboni[2], Dawid Szymanowski[3,4], Blair Schoene[4], Ingo Leya[5], Kevin D. McKeegan[6]

[1]Origins Laboratory, Department of the Geophysical Sciences and Enrico Fermi Institute, The University of Chicago, Chicago IL 60637, USA.
[2]School of Earth and Space Exploration, Arizona State University, Tempe, AZ 85281, USA.
[3]Institute of Geochemistry and Petrology, ETH Zurich, 8092 Zurich, Switzerland.
[4]Department of Geosciences, Princeton University, Princeton, NJ 08544, USA.
[5]Space Sciences and Planetology, University of Bern, Bern 3012, Switzerland.
[6]Department of Earth, Planetary, and Space Sciences, University of California, Los Angeles, CA 90095, USA.

[*]To whom correspondence should be addressed (dauphas@uchicago.edu)

**Abstract**

Crystallization of the lunar magma ocean yielded a chemically unique liquid residuum named KREEP. This component is expressed as a large patch on the near side of the Moon, and a possible smaller patch in the northwest portion of the Moon's South Pole-Aitken basin on the far side. Thermal models estimate that the crystallization of the lunar magma ocean (LMO) could have spanned from 10 and 200 Myr, while studies of radioactive decay systems have yielded inconsistent ages for the completion of LMO crystallization covering over 160 Myr. Here, we show that the Moon achieved >99% crystallization at 4429±76 Myr, indicating a lunar formation age of ~4450 Myr or possibly older. Using the $^{176}$Lu-$^{176}$Hf decay system ($t_{1/2}$=37 Gyr), we found that the initial $^{176}$Hf/$^{177}$Hf ratios of lunar zircons with varied U-Pb ages are consistent with their crystallization from a KREEP-rich reservoir with a consistently low $^{176}$Lu/$^{177}$Hf ratio of 0.0167 that emerged ~140 Myr after solar system formation. The previously proposed younger model age of ~4.33 Ga for the source of mare basalts (240 Myr after solar system formation) might reflect the timing of a large impact. Our results demonstrate that lunar magma ocean crystallization took place while the Moon was still battered by planetary embryos and planetesimals leftover from the main stage of planetary accretion. Study of Lu-Hf model ages for samples brought back from the South Pole-Aitken basin will help to assess the lateral continuity of KREEP and further understand its significance in the early history of the Moon.

**Significance Statement**

The Moon started as a fully molten body that gradually separated into layers as it cooled and crystallized. After 99% of the lunar magma ocean solidified, a unique residual liquid called KREEP, enriched in potassium (K), rare earth elements (REE), and phosphorus (P), was formed. Our study indicates that this KREEP liquid formed 4429±76 million years ago, approximately 140 million years after the solar system's birth. We also found that the KREEP liquid, as sampled by the Apollo missions, was remarkably uniform. Further studies of samples from the South Pole-Aitken basin will help clarify whether this uniformity extends laterally from the nearside to the farside of the Moon.

The mode and pace of Earth's growth are topics of considerable discussion, with two endmember theories involving fast accretion of small pebbles (1) or protracted accretion of large embryos thousands of kilometers in size (2). Where all models agree is that late in its history, the proto-Earth experienced one or several collisions with large planetary objects. One such impactor named Theia is speculated to have produced the Moon (3-5). Despite sustained efforts over decades to study samples brought back from the Moon by the Apollo, Luna, and Chang'E 5 missions, there is still considerable uncertainty on when the giant Moon-forming impact occurred (6-13). This impact could have been the last globally sterilizing event, and Earth might have been continuously habitable since then or shortly thereafter (14).

Geochemical evidence indicates that the Moon went through a magma ocean stage, whereby most or all of it was molten (15). Its cooling was associated with crystallization of a series of minerals with distinctive compositions, which drove the residual liquid to evolve chemically towards a composition called KREEP that is enriched in highly incompatible elements, notably potassium (K), rare earth elements (REEs), and phosphorus (P). KREEP was discovered in basalts recovered by the Apollo mission (15), and it was later detected remotely through γ-ray spectroscopy as two large patches of K and Th-enriched rocks positioned antipodally on the Moon in the Procellarum KREEP Terrane (PKT) and the South Pole-Aitken Terrane (SPAT) (16-20). Several strategies have been devised to date the formation of the KREEP reservoir, but no consensus has been reached on its age, with values spanning 160 Myr from 4.51 to 4.35 Ga (6-11). A robust age for KREEP would provide a minimum age for the Moon-forming impact.

The $^{176}$Lu-$^{176}$Hf decay system ($t_{1/2}$=37 Gyr (21, 22)) can be used to date the end of lunar magma ocean crystallization (7, 23-25). Application of this tool relies on the fact that during differentiation of the lunar magma ocean, Lu was preferentially retained in the mantle, while the crust and the residual melt layer known as KREEP became relatively enriched in Hf. As a result, once KREEP formed, its $^{176}$Hf/$^{177}$Hf ratio increased more slowly than the bulk Moon, which is assumed to be like chondritic meteorites (CHUR=Chondritic Uniform Reservoir)(26) because both Lu and Hf are refractory lithophile elements. By analyzing the isotopic compositions of bulk rocks and zircon minerals, one can determine when KREEP evolved as an isolated reservoir from CHUR and thus date KREEP formation. Since KREEP is thought to have formed when the LMO was 99% crystallized (27), dating the formation of this reservoir is equivalent to determining when LMO crystallization was nearly complete. Bulk rocks have been used for that purpose but these samples formed relatively late, requiring large extrapolation of the KREEP value backwards in time, which can lead to highly uncertain age estimates (23). A more robust approach is to measure the initial $^{176}$Hf/$^{177}$Hf (or ε$^{176}$Hf; the relative departure in part per $10^4$ from the CHUR ratio) of zircons that crystallized from lunar rocks containing a large KREEP component (7, 24, 25). Zircons are chemically resistant, have low Lu/Hf ratios, and their ages can be precisely determined using U-Pb geochronology, so they represent ideal time capsules to track the temporal evolution of ε$^{176}$Hf in the KREEP reservoir.

Taylor *et al.* (25) and Barboni *et al.* (7) analyzed lunar zircons using different methodologies (Fig. S4). While these studies provide valuable insights, their Lu-Hf data had significant uncertainties due to the use of peak stripping to correct for isobaric interferences during data reduction. To better define the age of KREEP, we have measured a new set of lunar zircons using improved methodologies (24) (see SI for details). These advances include separating Hf from elements that can cause isobaric interferences and accounting for the effect of Lu/Hf fractionation during sample processing. We have measured U/Pb ages, Hf isotopic compositions, and Lu/Hf ratios in lunar zircon leachates and residues treated by chemical abrasion intended to remove zircon domains more susceptible to Pb-loss or gain (24). The U/Pb ages were measured by isotope-dilution Thermal Ionization Mass Spectrometry at Princeton University. Hafnium from the same solutions was purified from interfering and all matrix elements including Zr for Hf isotopic analysis by MC-ICP-MS at the University of Chicago (24). The $^{176}Hf/^{177}Hf$ ratios were corrected for $^{176}Lu$-decay using measured Lu/Hf ratios and U-Pb ages, as well as for neutron capture effects associated with exposure to cosmic rays, using $\varepsilon^{178}Hf$ and $\varepsilon^{180}Hf$ as neutron dosimeters (23, 24) (Eq. S2). Some U-Pb ages and Hf isotopic analyses were previously reported, showing that a significant fraction of zircons crystallized in a short period of time starting at 4.338 Ga; a date that could correspond to the South Pole-Aitken impact (28). Indeed, this impact may have been powerful enough to cause the antipodal excavation of KREEP in the PKT (29). The full collection of zircons that we analyzed span 3.94 to 4.34 Ga in crystallization age, allowing us to constrain the age of KREEP and to evaluate the homogeneity of the Lu/Hf ratio in this reservoir. A potential difficulty with coupled U-Pb and Lu-Hf analyses of zircons is that Pb loss can occur without initial $\varepsilon^{176}Hf$ resetting. This can be remediated by chemical abrasion, which selectively removes domains susceptible to Pb loss, preserving closed-system domains that yield more concordant U-Pb ages (7, 24, 30). Most of our U-Pb ages are near-concordant and we found consistent $^{206}Pb$-$^{207}Pb$ ages between the different aliquots (L2 and R), indicating that the ages are most likely reliable (28), especially over the time span that we are interested in. The chemical abrasion procedure may introduce artifacts, most notably through the fractionation of the Lu/Hf ratio during dissolution if insoluble, Lu-rich fluorides precipitate (31). This could affect the correction of $\varepsilon^{176}Hf$ for in situ $^{176}Lu$ decay since zircon crystallization.

We measured a total of 36 zircon grains, and for many of these, we measured several fractions, corresponding to a total of 62 Lu-Hf and U-Pb analyses (Table S1, Datasets S1). A fraction of these were previously published to test the technique (24) and better understand the origin of the 4.338 Ga peak in the age distribution of lunar zircons (28). To evaluate the reliability of the data, we compare data for leachates and residues of chemical abrasion (24). For each zircon, three fractions were recovered during chemical abrasion. Leachate 1 (L1) was recovered after leaching with 90 µL 29 M HF for 6 h at 180°C. Leachate 2 (L2) was recovered after further processing the zircon through the same dissolution procedure. The residue (R) was finally dissolved in a Parr bomb using 90 µL 29 M HF for 60 h at 210°C. The first leachate was not used because it is prone to disturbance and contamination by common (non-radiogenic) Pb.

Our data are considered most reliable when L2 and R display similar ages and initial $\varepsilon^{176}Hf$ values, as this indicates that the zircon has a straightforward history, and that laboratory processing has not altered the intrinsic composition of the zircon (see Supplementary Information for detail). Indeed, any episode of partial Pb-loss or gain would have likely affected U-Pb ages of L2 and R differently, and any problem with data accuracy, correction of cosmogenic effects, or fractionation of Lu/Hf ratio during zircon dissolution would have resulted in different initial $\varepsilon^{176}Hf$ for L2 and R. The zircon measurements that yield consistent values between L2 and R are part of what we call *Tier 1*. There are 16 data points (initial $\varepsilon^{176}$Hf-U/Pb age) in this subset.

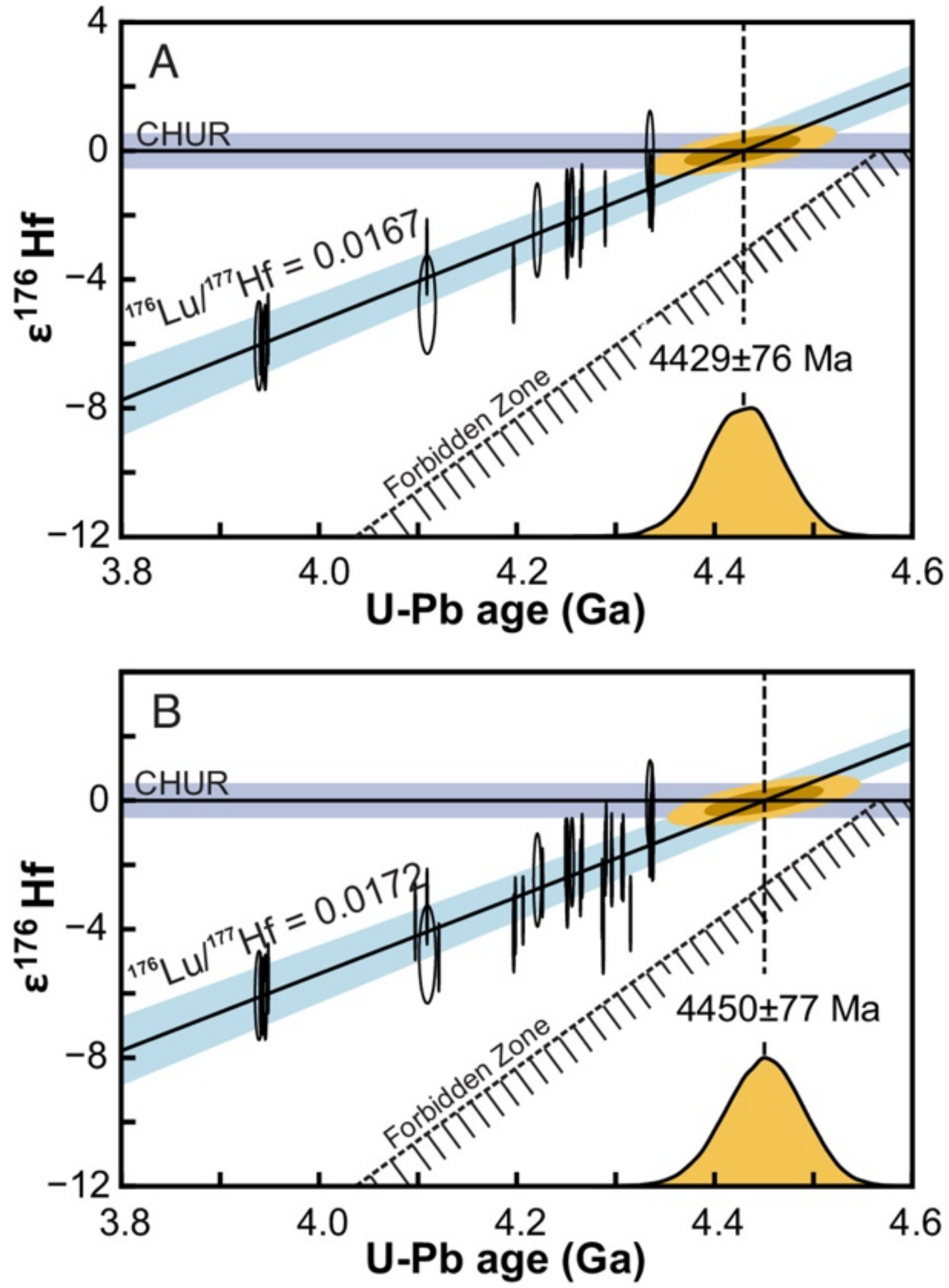

**Fig. 1.** Calculation of the $^{176}$Lu-$^{176}$Hf model age of KREEP magma formation based on lunar zircon $\varepsilon^{176}$Hf and U-Pb data (Table S1, Datasets S1). Panels A and B show the results for different tiers of data quality. *Tier 1* (A) corresponds to data where $\varepsilon^{176}$Hf values are consistent between leachate 2 (L2) and residue (R) of the chemical abrasion procedure, ensuring that the data are of the upmost quality ($n = 16$). *Tier 3* (B) comprises all R measurements, together with L2 measurements when they agree with R ($n = 51$). We focus on high quality *Tier 1* data (A), but all tiers (including intermediate *Tier 2*; see Supplementary Information) yield model ages and Lu/Hf ratios for KREEP that are identical within error. All data can be fit with a single line (MSWD=0.85 and 1.3 for *Tiers* 1 and 3, respectively), consistent with isolation of KREEP at a well-defined time (4429±76 Ma; the 68 and 95% confidence ellipses for the intercept with CHUR are shown in brown and yellow) and crystallization of the zircons at different times from melts of uniform $^{176}$Lu/$^{177}$Hf ratio (0.0167±0.0022). (±95% c.i.). The model age is given by the intercept between the best-fit line (solid black line with light blue 95% c.i.) and CHUR (horizonthal black solid line with dark blue 95% c.i). The $^{176}$Lu/$^{177}$Hf ratio of KREEP is given by the slope of the best-fit line (Eq. S10). The forbidden zone (hatched) corresponds to the minimum obtainable $\varepsilon^{176}$Hf value for a hypothetical reservoir formed at the formation of the solar system with $^{176}$Lu/$^{177}$Hf=0. The yellow curve on the x-axis is the marginal probability distribution for the model age of KREEP (the joint probability distribution is shown as an ellipse). All errors are 95% c.i. The $\varepsilon^{176}$Hf values shown here were corrected for cosmogenic effects using $\varepsilon^{178}$Hf (23, 24). Correcting these effects using $\varepsilon^{180}$Hf yields more scattered $\varepsilon^{176}$Hf values and more uncertain fit parameters (KREEP age= 4448±81 Ma, $^{176}$Lu/$^{177}$Hf=0.0162±0.0026 for *Tier 1*) that still overlap within error with $\varepsilon^{178}$Hf-corrected data (Fig. S6).

In most instances where the initial corrected $\varepsilon^{176}$Hf values differ between L2 and R, the raw $\varepsilon^{176}$Hf values agree. This discrepancy is often due to significant and varying Lu/Hf ratios (Fig. S2). It is highly unlikely for different domains within a single zircon to have originated with distinct initial $\varepsilon^{176}$Hf values and Lu/Hf ratios, and then fortuitously converge to similar present-day $\varepsilon^{176}$Hf values after $^{176}$Lu decay. The different initial corrected $\varepsilon^{176}$Hf values are most likely an analytical artifact from fractionation of the Lu/Hf during processing. The residues (R) hold the majority of Lu and Hf, and are therefore more reliable than L2. The subset of data that comprises all residue measurements (R) together with L2 measurements when they agree with R (*Tier 1* above) is called *Tier 3*. There are 40 data points in this subset. We also defined a *Tier 2* data set comprising 26 data that are intermediate in terms of reliability between the *Tier 1* and *Tier 3* data sets. The results are consistent with the other data sets and they are only discussed in Supplementary Information. Insoluble fluoride may be causing discrepancies between L2 and R in some samples. After leaching, we pipette out the leachate and rinse the residue multiple times with different acids, pipetting out each time. We then place the zircon residue back on the hotplate in 6 M HCl for at least 6 hours, followed by additional rinses with various acids. This process aims to dissolve fluorides; however, it may not have been fully effective. Further work will be necessary to assess if this issue impacts the results and, if so, to develop a mitigation strategy to achieve a higher proportion of *Tier 1* data.

The relationships between the $\varepsilon^{176}$Hf initial values and Pb-Pb ages are plotted in Fig. 1 for *Tiers 1* and *3* zircon datasets. The data can be fit with a single line (the reduced-$\chi^2$ also known as Mean Square Weighted Deviation MSWD are 0.85 and 1.3 for *Tiers 1* and *3* datasets with $n-2=16$ and 38 degrees of freedom, respectively; the 2.5-97.5% interquantile range for the reduced-$\chi^2$ distribution for those degrees of freedom are 0.43-1.80 and 0.60-1.50), and are therefore consistent with all zircons crystallizing at different times from a melt of uniform Hf isotopic composition. The intercepts between the best fit lines and the $x$-axes give model ages for KREEP of 4429±76 Ma (±95% c.i.) and 4450±77 (±95% c.i.) Myr for *Tiers 1* and *3*, respectively. The slopes of the $\varepsilon^{176}$Hf-age regressions correspond to $^{176}$Lu/$^{177}$Hf ratios of 0.0167±0.0022 (±95% c.i.) and 0.0172±0.0016 (±95% c.i.) for *Tiers 1* and *3*, respectively (Eq. S10). There is good agreement between all tiers of data quality, indicating that the results are not influenced by our parsing in tiers. We use *Tier 1* results for discussion as they are identical within error with *Tier 3* but are less likely to be affected by any form of bias.

The $^{176}$Lu/$^{177}$Hf ratio of KREEP inferred here is consistent but more precise than prior estimates obtained by measuring the trace element composition of KREEP-enriched rocks returned from the Moon by the Apollo mission, which gave ratios of 0.0164 (25), 0.0154±0.0034 (23), and 0.00153±0.0033 (32). This supports the view that zircons indeed crystallized from a relatively uniform KREEP reservoir. Taylor *et al.* (25) analyzed zircon grains *in situ* using Secondary Ion Mass Spectrometry for U-Pb and laser ablation multicollector inductively coupled plasma mass spectrometry (LA-MC-ICP-MS) for Lu-Hf, and found significant scatter in the data beyond individual data uncertainty (Fig. S4), with a peak in the model age distribution at ~4.48 Ga. Barboni *et al.* (7) also found significant scatter (Fig. S4) with model age estimates for individual zircons that largely overlap with ours, but with a few data points giving older model ages. The most critical data that give older ages have large uncertainties, on which the present study improves. The present Hf isotopic data have higher precision, and their accuracy is improved through purification of Hf by chromatography.

Borg and Carlson (9) made a case for crystallization of much of the lunar magma ocean at 4.33 Ga, with the strongest piece of evidence provided by $^{146}$Sm-$^{142}$Nd ($t_{1/2}$=103 Myr) and $^{147}$Sm-$^{143}$Nd ($t_{1/2}$=106 Gyr) systematics applied to the source of mare basalts. We have re-examined the data set of Borg *et al*. (8) and agree with their assessment that a model age of lunar magma ocean differentiation of 4.44 Ga provides a poorer fit to initial $^{142}$Nd/$^{144}$Nd ratios of mare basalts compared to a model age of 4.33 Ga (Fig. S5), but in both cases, there is significant scatter with 5 out of 30 samples that cannot be explained. A difficulty with the interpretation of LMO crystallization at 4.33 Ga is that single lunar zircon mineral grains have been dated at 4.42 Ga (6), contradicting the view that LMO differentiation occurred late. Older zircon grains have also been found on Earth (33, 34). Interestingly, the 4.33 Ga age inferred for the source of the mare basalts also corresponds to a marked and narrow peak (~4 Myr duration) in the age distribution of U-Pb zircon ages, which Barboni *et al*. (28) interpreted to correspond to large scale melting induced by a large impact, possibly associated with formation of the South Pole-Aitken basin. Even if such an impact did not induce complete melting of the refractory ultramafic cumulate that is thought to be the main source of the mare basalts, it might have induced melting of the most fusible components of the cumulate (35), allowing Sm and Nd redistribution and equilibration through reactive melt infiltration (36). It could have also induced mixing and subsequent density separation of minerals within the Moon (37). Both factors could have contributed to resetting Sm-Nd systematics at a bulk rock scale, so the Sm-Nd model age may date a late-stage large impact rather than early crystallization of the lunar magma ocean.

We obtain an age for separation of KREEP of 4429±76 Ma (Fig. 1). The oldest zircon that we have analyzed here is 4338 Ma, but previous studies have reported older single mineral ages that overlap with our KREEP model age. Nemchin *et al*. (6) reported an age of 4417±6 Ma in a zircon from a lunar breccia using an *in situ* technique. Zhang *et al*. (10) also reported an age of 4460±30 Ma in a zircon. The authors originally dismissed the data because applying another technique on the same zircon yielded an age 300 Ma younger. However, Greer *et al*. (11) found no evidence for secondary disturbance in this zircon and argued that the older age was real. Old ages (>4.4 Ga) were also reported for some ferroan anorthosites (38-41) that are thought to represent flotation of a plagioclase crust during LMO crystallization. These older ages were dismissed due to the lack of concordance among different radiochronometers and disagreement with more recent data (9). Borg and Carlson (9) argued that the formation of the ferroan anorthosite suite was most reliably dated using Sm-Nd systematics applied to Apollo samples 60025 (37), 62237 (42), and Y-86032 (39), with sample 60025 yielding the most precise age estimate of 4.367±0.011 Ga. However, the Sm-Nd crystallization ages obtained from this sample do not all agree, as Carlson and Lugmair (40) reported a notably older age of 4.44±0.02 Ga. Establishing the concordance with Pb-Pb ages is challenging, given that plagioclase and pyroxene yield disparate ages in this sample (37, 43). Sample 60025 is a polymict breccia containing materials not all derived from the LMO (44). The age discrepancy may therefore stem from differences in the analyzed materials, with the older age of 4.44±0.02 Ga (40) potentially representing the formation of a flotation crust.

The Lu-Hf model age of KREEP formation corresponds to the time when the LMO reached 99 to 99.9% crystallization (Fig. S6) (27). The age of KREEP therefore gives a minimum age for the Moon. To go beyond this and provide a solid constraint on the time of the giant impact, one must rely on uncertain models of lunar magma ocean cooling. A few thousands of years is all it took for the lunar magma ocean to cool to 80% crystallization (45). What happened beyond this is uncertain. Formation of a plagioclase flotation crust likely hampered further heat loss. In addition, some heat was deposited inside the Moon by tidal dissipation (45-49). Models involving a simple stagnant conductive lid predict a cooling time of 10 to 30 Myr (45), meaning that the Moon could have formed 4449±76 Myr ago. Interestingly, this is close to the age of 4.44±0.02 Ga obtained for ferroan anorthosite 60025 by Carlson and Lugmair (40), possibly dating formation of a flotation crust. However, some models using low thermal conductivity for the anorthositic crust and considering extraction of melt from mafic cumulate predict a crystallization time of 200 Myr (46). Our results on the age of KREEP show that such a prolonged crystallization time is unlikely because it would put lunar formation before solar system formation, but a 150 Myr cooling time would agree with current knowledge. It has been argued, based on $^{182}$Hf-$^{182}$W systematics, that the Moon could have formed 40 to 74 Myr after the birth of the solar system (4.53 to 4.49 Ga) (50, 51). However, this age may lack significance if the $^{182}$W excess in lunar rocks relative to terrestrial rocks is due to disproportionate late accretion of meteoritic material after core formation rather than $^{182}$Hf-decay (13). Taken at face value, such an age for lunar formation would mean that the LMO took ~60 Myr to be fully solidified.

This study shows that zircons recovered from the Moon by the Apollo missions were all derived from a single KREEP reservoir that was isolated from the bulk silicate Moon at 4.43 Ga. Remote sensing γ-ray mapping of K and Th on the Moon has revealed the extent of the Procellarum KREEP Terrane (18, 52) that is likely the predominant chemical source of the lunar zircons found in rocks from the Apollo missions. The same mapping showed that a smaller patch of material enriched in KREEP is also present in the South Pole-Aitken terrane, on the antipode of the Procellarum KREEP Terrane (19, 20). The Chang'e 6 mission retrieved rocks from the South Pole-Aitken basin. If zircons are found in these rocks, they should have experienced a different impact history than those from the Apollo missions, with the South Pole-Aitken impact expected to be featured predominantly in their age distribution. If KREEP was a uniform layer, which is the prevailing view, we expect that those zircons will plot on the same $\varepsilon^{176}$Hf-age trend as the one established here, regardless of differences in their age distribution. But the alternative that KREEP, which formed when the Moon was 99% crystallized, was in fact composed of pools of magmas isolated at different times cannot be excluded. As we enter a new era of Moon exploration and sampling, our determination of the age of KREEP serves as a fundamental reference for testing hypotheses regarding its nature and occurrence within the South Pole-Aitken basin.

**Acknowledgment.** The support of the late Reika Yokochi through the years was instrumental to the success of the Origins Laboratory. We thank two reviewers for their constructive critiques, which significantly improved this manuscript. ND thanks Shogo Tachibana, Tomohiro Usui, Tsuyoshi Iizuka, Tetsuya Yokoyama, and the staff of the Department of Earth and Planetary Science at the University of Tokyo for their hospitality and enriching scientific discussions during a visit in the summer of 2024. This work was supported by grants 80NSSC20K0821 (NASA-EW) to KM, MB, ND, and BS, and by grants 80NSSC21K0380 (NASA-EW), 80NSSC20K1409 (NASA-HW), EAR-2001098 (NSF-CSEDI), DE-SC0022451 (DOE), 80NSSC23K1022 (NASA-LARS), 80NSSC23K1163 (NASA-MMX) to ND.

## References

1. A. Johansen *et al.*, A pebble accretion model for the formation of the terrestrial planets in the Solar System. *Science advances* **7**, eabc0444 (2021).
2. J. Chambers, G. Wetherill, Making the terrestrial planets: N-body integrations of planetary embryos in three dimensions. *Icarus* **136**, 304-327 (1998).
3. W. K. Hartmann, D. R. Davis, Satellite-sized planetesimals and lunar origin. *Icarus* **24**, 504-515 (1975).
4. A. G. W. Cameron, W. R. Ward, The Origin of the Moon. *Lunar and Planetary Science Conference* **7**, 120 (1976).
5. R. M. Canup *et al.*, Origin of the Moon. *Reviews in Mineralogy and Geochemistry* **89**, 53-102 (2023).
6. A. Nemchin *et al.*, Timing of crystallization of the lunar magma ocean constrained by the oldest zircon. *Nature geoscience* **2**, 133-136 (2009).
7. M. Barboni *et al.*, Early formation of the Moon 4.51 billion years ago. *Science advances* **3**, e1602365 (2017).
8. L. E. Borg *et al.*, Isotopic evidence for a young lunar magma ocean. *Earth and Planetary Science Letters* **523**, 115706 (2019).
9. L. E. Borg, R. W. Carlson, The Evolving Chronology of Moon Formation. *Annual Review of Earth and Planetary Sciences* **51**, 25-52 (2023).
10. B. Zhang *et al.*, Radiogenic Pb mobilization induced by shock metamorphism of zircons in the Apollo 72255 Civet Cat norite clast. *Geochimica et Cosmochimica Acta* **302**, 175-192 (2021).
11. J. Greer *et al.*, 4.46 Ga zircons anchor chronology of lunar magma ocean. *Geochemical Perspectives Letters* **27**, 49-53 (2023).
12. M. M. Thiemens, P. Sprung, R. O. Fonseca, F. P. Leitzke, C. Münker, Early Moon formation inferred from hafnium–tungsten systematics. *Nature Geoscience* **12**, 696-700 (2019).
13. T. S. Kruijer, G. J. Archer, T. Kleine, No 182W evidence for early Moon formation. *Nature Geoscience* **14**, 714-715 (2021).
14. R. I. Citron, S. T. Stewart, Large impacts onto the early Earth: planetary sterilization and iron delivery. *The Planetary Science Journal* **3**, 116 (2022).
15. P. H. Warren, The magma ocean concept and lunar evolution. *IN: Annual review of earth and planetary sciences. Volume 13. Palo Alto, CA, Annual Reviews, Inc., 1985, p. 201-240.* **13**, 201-240 (1985).
16. D. Lawrence *et al.*, Small-area thorium features on the lunar surface. *Journal of Geophysical Research: Planets* **108** (2003).
17. T. H. Prettyman *et al.*, Elemental composition of the lunar surface: Analysis of gamma ray spectroscopy data from Lunar Prospector. *Journal of Geophysical Research: Planets* **111** (2006).
18. S. Kobayashi *et al.*, Lunar farside Th distribution measured by Kaguya gamma-ray spectrometer. *Earth and Planetary Science Letters* **337**, 10-16 (2012).

19. M. Naito *et al.*, Potassium and thorium abundances at the South Pole‐Aitken basin obtained by the Kaguya gamma‐ray spectrometer. *Journal of Geophysical Research: Planets* **124**, 2347-2358 (2019).
20. J. J. Hagerty, D. Lawrence, B. Hawke, Thorium abundances of basalt ponds in South Pole‐Aitken basin: Insights into the composition and evolution of the far side lunar mantle. *Journal of Geophysical Research: Planets* **116** (2011).
21. U. Söderlund, P. J. Patchett, J. D. Vervoort, C. E. Isachsen, The 176Lu decay constant determined by Lu–Hf and U–Pb isotope systematics of Precambrian mafic intrusions. *Earth and Planetary Science Letters* **219**, 311-324 (2004).
22. T. Hayakawa, T. Shizuma, T. Iizuka, Half-life of the nuclear cosmochronometer 176Lu measured with a windowless 4 π solid angle scintillation detector. *Communications Physics* **6**, 299 (2023).
23. P. Sprung, T. Kleine, E. E. Scherer, Isotopic evidence for chondritic Lu/Hf and Sm/Nd of the Moon. *Earth and Planetary Science Letters* **380**, 77-87 (2013).
24. X. Chen *et al.*, Methodologies for 176Lu–176Hf Analysis of Zircon Grains from the Moon and Beyond. *ACS Earth and Space Chemistry* 10.1021/acsearthspacechem.3c00093 (2023).
25. D. J. Taylor, K. D. McKeegan, T. M. Harrison, Lu–Hf zircon evidence for rapid lunar differentiation. *Earth and Planetary Science Letters* **279**, 157-164 (2009).
26. A. Bouvier, J. D. Vervoort, P. J. Patchett, The Lu–Hf and Sm–Nd isotopic composition of CHUR: constraints from unequilibrated chondrites and implications for the bulk composition of terrestrial planets. *Earth and Planetary Science Letters* **273**, 48-57 (2008).
27. P. H. Warren, J. T. Wasson, The origin of KREEP. *Reviews of Geophysics* **17**, 73-88 (1979).
28. M. Barboni *et al.*, High precision U-Pb zircon dating identifies a major magmatic event on the Moon at 4.338 Ga. *Science Advances* **10**, eadn9871 (2024).
29. M. J. Jones *et al.*, A South Pole–Aitken impact origin of the lunar compositional asymmetry. *Science Advances* **8**, eabm8475 (2022).
30. J. M. Mattinson, Zircon U–Pb chemical abrasion ("CA-TIMS") method: combined annealing and multi-step partial dissolution analysis for improved precision and accuracy of zircon ages. *Chemical Geology* **220**, 47-66 (2005).
31. J. Blichert-Toft, F. Albarède, The Lu-Hf isotope geochemistry of chondrites and the evolution of the mantle-crust system. *Earth and Planetary Science Letters* **148**, 243-258 (1997).
32. A. M. Gaffney, L. E. Borg, A young solidification age for the lunar magma ocean. *Geochimica et Cosmochimica Acta* **140**, 227-240 (2014).
33. S. A. Wilde, J. W. Valley, W. H. Peck, C. M. Graham, Evidence from detrital zircons for the existence of continental crust and oceans on the Earth 4.4 Gyr ago. *Nature* **409**, 175-178 (2001).
34. P. Holden *et al.*, Mass-spectrometric mining of Hadean zircons by automated SHRIMP multi-collector and single-collector U/Pb zircon age dating: The first 100,000 grains. *International Journal of Mass Spectrometry* **286**, 53-63 (2009).

35. C. L. McLeod, A. D. Brandon, R. M. Armytage, Constraints on the formation age and evolution of the Moon from 142Nd–143Nd systematics of Apollo 12 basalts. *Earth and Planetary Science Letters* **396**, 179-189 (2014).
36. G. Borghini *et al.*, Fast REE re-distribution in mantle clinopyroxene via reactive melt infiltration. *Geochemical Perspectives Letters* **26**, 40-44 (2023).
37. L. E. Borg, J. N. Connelly, M. Boyet, R. W. Carlson, Chronological evidence that the Moon is either young or did not have a global magma ocean. *Nature* **477**, 70-72 (2011).
38. M. D. Norman, L. E. Borg, L. E. Nyquist, D. D. Bogard, Chronology, geochemistry, and petrology of a ferroan noritic anorthosite clast from Descartes breccia 67215: Clues to the age, origin, structure, and impact history of the lunar crust. *Meteoritics & Planetary Science* **38**, 645-661 (2003).
39. L. Nyquist *et al.*, Feldspathic clasts in Yamato-86032: Remnants of the lunar crust with implications for its formation and impact history. *Geochimica et Cosmochimica Acta* **70**, 5990-6015 (2006).
40. R. W. Carlson, G. W. Lugmair, The age of ferroan anorthosite 60025: oldest crust on a young Moon? *Earth and Planetary Science Letters* **90**, 119-130 (1988).
41. C. Alibert, M. D. Norman, M. T. McCulloch, An ancient Sm-Nd age for a ferroan noritic anorthosite clast from lunar breccia 67016. *Geochimica et Cosmochimica Acta* **58**, 2921-2926 (1994).
42. C. Sio, L. Borg, W. Cassata, The timing of lunar solidification and mantle overturn recorded in ferroan anorthosite 62237. *Earth and Planetary Science Letters* **538**, 116219 (2020).
43. B. Hanan, G. Tilton, 60025: relict of primitive lunar crust? *Earth and Planetary Science Letters* **84**, 15-21 (1987).
44. M. Torcivia, C. Neal, Unraveling the components within Apollo 16 ferroan anorthosite suite cataclastic anorthosite sample 60025: Implications for the lunar magma ocean model. *Journal of Geophysical Research: Planets* **127**, e2020JE006799 (2022).
45. L. T. Elkins-Tanton, S. Burgess, Q.-Z. J. E. Yin, P. S. Letters, The lunar magma ocean: Reconciling the solidification process with lunar petrology and geochronology. **304**, 326-336 (2011).
46. M. Maurice, N. Tosi, S. Schwinger, D. Breuer, T. Kleine, A long-lived magma ocean on a young Moon. *Science advances* **6**, eaba8949 (2020).
47. V. Perera, A. P. Jackson, L. T. Elkins-Tanton, E. Asphaug, Effect of reimpacting debris on the solidification of the lunar magma ocean. *Journal of Geophysical Research: Planets* **123**, 1168-1191 (2018).
48. J. Meyer, L. Elkins-Tanton, J. Wisdom, Coupled thermal–orbital evolution of the early Moon. *Icarus* **208**, 1-10 (2010).
49. Z. Tian, J. Wisdom, L. Elkins-Tanton, Coupled orbital-thermal evolution of the early Earth-Moon system with a fast-spinning Earth. *Icarus* **281**, 90-102 (2017).
50. M. M. Thiemens *et al.*, Reply to: No 182W evidence for early Moon formation. *Nature Geoscience* **14**, 716-718 (2021).

51. M. M. Thiemens, P. Sprung, R. O. C. Fonseca, F. P. Leitzke, C. Münker, Early Moon formation inferred from hafnium–tungsten systematics. *Nature Geoscience* **12**, 696-700 (2019).
52. D. Lawrence *et al.*, Global elemental maps of the Moon: The Lunar Prospector gamma-ray spectrometer. *Science* **281**, 1484-1489 (1998).
53. M. Barboni *et al.*, High precision U-Pb zircon dating pinpoints a major magmatic event on the Moon at 4.337 Ga. *Science Advances* **10**, eadn9871 (2024).
54. C. Meyer, I. S. Williams, W. Compston, Uranium-lead ages for lunar zircons: Evidence for a prolonged period of granophyre formation from 4.32 to 3.88 Ga. *Meteoritics & Planetary Science* **31**, 370-387 (1996).
55. C. Meyer, Lunar sample compendium. (2005).
56. L. A. Haskin, R. L. Korotev, J. J. Gillis, B. L. Jolliff (2002) Stratigraphies of Apollo and Luna highland landing sites and provenances of materials from the perspective of basin impact ejecta modeling. in *Lunar and Planetary Science Conference*, p 1364.
57. R. Morrison, V. Oberbeck (1975) Geomorphology of crater and basin deposits-Emplacement of the Fra Mauro formation. in *In: Lunar Science Conference, 6th, Houston, Tex., March 17-21, 1975, Proceedings. Volume 3.(A78-46741 21-91) New York, Pergamon Press, Inc., 1975, p. 2503-2530.*, pp 2503-2530.
58. F. J. Stadermann, E. Heusser, E. K. Jessberger, S. Lingner, D. Stöffler, The case for a younger Imbrium basin: New 40Ar-39Ar ages of Apollo 14 rocks. *Geochimica et Cosmochimica Acta* **55**, 2339-2349 (1991).
59. R. Drozd, C. Hohenberg, C. Morgan, C. Ralston, Cosmic-ray exposure history at the Apollo 16 and other lunar sites: lunar surface dynamics. *Geochimica et Cosmochimica Acta* **38**, 1625-1642 (1974).
60. R. E. Merle *et al.*, Origin and transportation history of lunar breccia 14311. *Meteoritics & Planetary Science* **52**, 842-858 (2017).
61. M. Hopkins, S. J. Mojzsis, A protracted timeline for lunar bombardment from mineral chemistry, Ti thermometry and U–Pb geochronology of Apollo 14 melt breccia zircons. *Contributions to Mineralogy and Petrology* **169**, 1-18 (2015).
62. J. M. Devine, D. S. McKay, J. J. Papike, Lunar regolith: Petrology of the< 10 µm fraction. *Journal of Geophysical Research: Solid Earth* **87**, A260-A268 (1982).
63. T. Labotka, M. Kempa, C. White, J. Papike, J. Laul (1980) The lunar regolith-Comparative petrology of the Apollo sites. in *In: Lunar and Planetary Science Conference, 11th, Houston, TX, March 17-21, 1980, Proceedings. Volume 2.(A82-22296 09-91) New York, Pergamon Press, 1980, p. 1285-1305.*, pp 1285-1305.
64. C. A. Crow, K. D. McKeegan, D. E. Moser, Coordinated U–Pb geochronology, trace element, Ti-in-zircon thermometry and microstructural analysis of Apollo zircons. *Geochimica et Cosmochimica Acta* **202**, 264-284 (2017).
65. M. Grange, A. Nemchin, R. Pidgeon, The effect of 1.9 and 1.4 Ga impact events on 4.3 Ga zircon and phosphate from an Apollo 15 melt breccia. *Journal of Geophysical Research: Planets* **118**, 2180-2197 (2013).
66. P. H. Warren, G. J. Taylor, K. Keil, D. N. Shirley, J. T. Wasson, Petrology and chemistry of two "large" granite clasts from the Moon. *Earth and Planetary Science Letters* **64**, 175-185 (1983).

67. A. Nemchin, R. Pidgeon, M. Whitehouse, J. P. Vaughan, C. Meyer, SIMS U–Pb study of zircon from Apollo 14 and 17 breccias: implications for the evolution of lunar KREEP. *Geochimica et Cosmochimica Acta* **72**, 668-689 (2008).
68. A. Nemchin, M. Whitehouse, R. Pidgeon, C. Meyer, Oxygen isotopic signature of 4.4–3.9 Ga zircons as a monitor of differentiation processes on the Moon. *Geochimica et Cosmochimica Acta* **70**, 1864-1872 (2006).
69. M. Grange, R. Pidgeon, A. Nemchin, N. E. Timms, C. Meyer, Interpreting U–Pb data from primary and secondary features in lunar zircon. *Geochimica et Cosmochimica Acta* **101**, 112-132 (2013).
70. A. Pourmand, N. Dauphas, Distribution coefficients of 60 elements on TODGA resin: application to Ca, Lu, Hf, U and Th isotope geochemistry. *Talanta* **81**, 741-753 (2010).
71. J. Zhang, N. Dauphas, A. M. Davis, A. Pourmand, A new method for MC-ICPMS measurement of titanium isotopic composition: Identification of correlated isotope anomalies in meteorites. *Journal of Analytical Atomic Spectrometry* **26**, 2197-2205 (2011).
72. C. Münker, S. Weyer, E. Scherer, K. Mezger, Separation of high field strength elements (Nb, Ta, Zr, Hf) and Lu from rock samples for MC-ICPMS measurements. *Geochemistry, Geophysics, Geosystems* **2** (2001).
73. T. Iizuka, T. Yamaguchi, Y. Hibiya, Y. Amelin, Meteorite zircon constraints on the bulk Lu− Hf isotope composition and early differentiation of the Earth. *Proceedings of the National Academy of Sciences* **112**, 5331-5336 (2015).
74. H. Gerstenberger, G. Haase, A highly effective emitter substance for mass spectrometric Pb isotope ratio determinations. *Chemical geology* **136**, 309-312 (1997).
75. D. Szymanowski, B. Schoene, U–Pb ID-TIMS geochronology using ATONA amplifiers. *Journal of Analytical Atomic Spectrometry* **35**, 1207-1216 (2020).
76. J. N. Connelly *et al.*, The absolute chronology and thermal processing of solids in the solar protoplanetary disk. *Science* **338**, 651-655 (2012).
77. T. Harrison *et al.*, Heterogeneous Hadean hafnium: evidence of continental crust at 4.4 to 4.5 Ga. *Science* **310**, 1947-1950 (2005).
78. N. Marks, L. Borg, I. Hutcheon, B. Jacobsen, R. Clayton, Samariumneodymium chronology of an Allende calcium-aluminum-rich inclusion with implications for 146Sm isotopic evolution. *Earth Planet. Sci. Lett* **405**, 15-24 (2014).
79. L. E. Borg, A. M. Gaffney, C. K. Shearer, A review of lunar chronology revealing a preponderance of 4.34–4.37 Ga ages. *Meteoritics & Planetary Science* **50**, 715-732 (2015).
80. Y. Amelin, E. Rotenberg, Sm–Nd systematics of chondrites. *Earth and Planetary Science Letters* **223**, 267-282 (2004).
81. L. Fang *et al.*, Half-life and initial Solar System abundance of 146Sm determined from the oldest andesitic meteorite. *Proceedings of the National Academy of Sciences* **119**, e2120933119 (2022).
82. C. Burkhardt *et al.*, A nucleosynthetic origin for the Earth's anomalous 142Nd composition. *Nature* **537**, 394-398 (2016).

83. L. Nyquist *et al.*, 146Sm-142Nd formation interval for the lunar mantle. *Geochimica et Cosmochimica Acta* **59**, 2817-2837 (1995).
84. M. Boyet, R. W. Carlson, A highly depleted moon or a non-magma ocean origin for the lunar crust? *Earth and Planetary Science Letters* **262**, 505-516 (2007).
85. A. D. Brandon *et al.*, Re-evaluating 142Nd/144Nd in lunar mare basalts with implications for the early evolution and bulk Sm/Nd of the Moon. *Geochimica et Cosmochimica Acta* **73**, 6421-6445 (2009).
86. J. E. Dickinson Jr, P. Hess, Zircon saturation in lunar basalts and granites. *Earth and Planetary Science Letters* **57**, 336-344 (1982).
87. G. A. Gualda, M. S. Ghiorso, R. V. Lemons, T. L. Carley, Rhyolite-MELTS: a modified calibration of MELTS optimized for silica-rich, fluid-bearing magmatic systems. *Journal of Petrology* **53**, 875-890 (2012).
88. L. J. Crisp, A. J. Berry, A new model for zircon saturation in silicate melts. *Contributions to Mineralogy and Petrology* **177**, 71 (2022).

# Supporting Information for

# Completion of Lunar Magma Ocean Solidification at 4.43 Ga

Nicolas Dauphas[1*], Zhe J. Zhang[1], Xi Chen[1], Mélanie Barboni[2], Dawid Szymanowski[3,4], Blair Schoene[4], Ingo Leya[5], Kevin D. McKeegan[6]

[1]Origins Laboratory, Department of the Geophysical Sciences and Enrico Fermi Institute, The University of Chicago, Chicago IL 60637, USA.
[2]School of Earth and Space Exploration, Arizona State University, Tempe, AZ 85281, USA.
[3]Institute of Geochemistry and Petrology, ETH Zurich, 8092 Zurich, Switzerland.
[4]Department of Geosciences, Princeton University, Princeton, NJ 08544, USA.
[5]Space Sciences and Planetology, University of Bern, Bern 3012, Switzerland.
[6]Department of Earth, Planetary, and Space Sciences, University of California, Los Angeles, CA 90095, USA.

Corresponding author: Nicolas Dauphas
Email: dauphas@uchicago.edu

**This PDF file includes:**

Supporting text
Figures S1 to S7
Tables S1
SI References

**Other supporting materials for this manuscript include the following:**

Datasets S1

## Supporting Information Text

Unless otherwise noted, error bars are 95% confidence intervals.

### 1. Samples

The analyzed zircon grains analyzed were extracted from Apollo samples by gently crushing the samples, separating dense minerals using heavy liquid methylene iodide, and handpicking zircon fragments from the high density (>3.3 g/cm$^3$) fraction under UV light (25). To maximize the zircon yield, heavy residues were then put on double-sided tape, put in a Tescan SEM coupled with EDS at UCLA and each grain was tested for zirconium. The zircons were mounted in epoxy and characterized using cathodoluminescence to avoid grains with obvious complex growth patterns. The zircons were then dated using U-Pb by Secondary Ion Mass Spectrometry (SIMS) to make sure that these were ancient lunar zircons suitable for Lu-Hf study. As discussed below, the ages of these zircons were re-measured using U-Pb Isotope Dilution-Thermal Ionization Mass Spectrometry (ID-TIMS), which is a more precise technique, so the SIMS U-Pb ages are only used for selecting old zircons. In total, 36 zircon grains were studied using the same protocol. Chen *et al*. (24) previously reported Hf isotopic data for 5 zircon grains (14163 z9, 14163 z89, 14163 z26, 14321 z3, 72275 z1), and Barboni *et al*. (53) reported Hf isotopic data for additional 6 zircon grains (14311,58 z7, 14311,58 z21, 14311,58 z37, 14163,65 z3, 14163,65 z7, 15405,75 z1). We report here data for 25 new zircon grains, and we use the data from the 11 reported zircon grains from Chen *et al*. (24) and Barboni *et al*. (53), as these zircons were analyzed using the exact same protocol, and dating KREEP crystallization was not the scope of those studies. Chen *et al*. (24) used 5 small zircon grains to test the methodology, while Barboni *et al*. (53) focused on the 4.338 Ga peak in the U-Pb age distribution, so they only report Hf data for zircon grains with U-Pb ages between 4334 and 4338 Ma. Zircons in igneous clasts are rare, and they are usually found as individual grains in breccia matrices or as detrital grains in soil samples (54). Little direct information is therefore lost by not knowing the petrographic context of their occurrences. The zircons were extracted from the following samples (as indicated in the zircon labels)(55):

- 26 zircon grains from Apollo 14 "14311". Sample 14311 is an impact melt breccia (3204 g) from a site covered by the Fra Mauro Formation, which is thought to contain ~15 to 60% ejecta from the Imbrium impact basin (56, 57). The matrix (75-95% of the total volume) is melted and recrystallized, reacting with the clasts that comprise various mineral (plagioclase, pyroxene, Fe-Ti oxides) and lithic (igneous rocks and older impact breccias) fragments (55). The exposure age of this breccia is ~550-660 Myr (58, 59). Zircons from this breccia have been extensively characterized for their U-Pb ages, REE patterns, and Ti-thermometry (53, 54, 60, 61). They are crystals and fragments predating breccia formation (60). These zircons yield concordant U-Pb ages spanning ~3.95 to ~4.35 Ga. The ages are unevenly distributed, defining peaks at 4.337, 4.240, 4.110, 4.030, and 3.960 Ga that could reflect crystallization from impact melt sheets associated with large impacts. The REE composition of the oldest zircon population is like that of granite and gabbronorite clasts. Barboni *et al*. (53) report Hf isotopic data for 3 of the 24 zircon grains from this sample (z9, z21, z37). We report data for 21 new zircon grains from this sample (z57, z8, z7, z18, z38, z43, z69, z71, z40, z15, z12, z47, z59, z61, z64, z6, z58, z60, z34, z24, and z27).
- 7 zircon grains from the Apollo 14 "14163" bulk soil sample (7776 g). It contains a large percentage of glass; most of it (46-61%) in the form of agglutinates. Smaller soil size fractions contain abundant granitic glass (62, 63). Zircons from that soil have U-Pb ages that are mostly concordant and span ~3.95 to 4.35 Ga (53, 64). Data for 3 out of those 7 zircons were presented in Chen *et al*. (24) (z9, z89, z26). Data for 2 additional zircon grains were presented in Barboni *et al*. (53) (14163_65 z3, 14163_65 z7). We report data for 2 new zircon grains from this sample (14163_65 z4 and 14163_949 z3).
- 1 zircon grain from Apollo 15 "15405". Sample 15405 is a breccia (513.1 g) containing mineral fragments as well as lithic clasts of KREEP-rich basalt, granite, and quartz monzodiorite. The matrix has a similar composition to KREEP-rich basalts. U-Pb dating of zircons indicates relatively recent Pb-loss presumably associated with formation of the breccia (54, 65), but concordant ages of up to 4.3-4.4 Ga are found (53, 54, 64, 65). Data for this zircon grain was presented in Barboni *et al*. (53) (14163_75 z1). No new data are reported here.

- 1 zircon grain from Apollo 14 "14321". Sample 14321 is a clast-rich crystalline matrix breccia (8998 g) sampled very near the Cone crater. The matrix is mostly crystalline. The clasts are diverse in composition, including basalt, troctolite, anorthosite, dunite, and granite. Granite clast 14321,1027 (66) contains a zircon dated at 3.965 Ga (54). Anorthosite clast 14321.16 contains a zircon dated at 4.02-4.05 Ga (67). More zircons without petrographic context have been dated using U-Pb and the ages span the range 3.9-4.4 Ga (53, 64, 67-69). Data for zircon grain z3 was reported in Chen *et al*. (24). No new data are reported here.
- 1 zircon grain from Apollo 17 72275A. Sample 72275 is a feldspathic polymict breccia (3640 g) with ~60% matrix and ~40% clasts. It may represent ejecta from the Serenitatis basin. Most zircons from this breccia have U-Pb ages of 4.24-4.37 Ga (6, 53, 64), with one zircon holding a record precise concordant age of 4,417±6 Ma (6), which represents a minimum age for lunar formation. Various clast types have been documented, including a coarse-grained granitic clast (72275,520; Fig. 1 of Meyer *et al*. 1996) (54). Data for 1 zircon grain was reported in Chen *et al*. (24), which was also compiled by Barboni *et al*. (53) (z1). No new data are reported here.

## 2. Sample processing and U-Pb-Lu-Hf isotopic analyses

Sample processing and isotopic analyses were explained in Chen *et al*. (24). That paper also contains an in-depth discussion of cosmogenic effects and their correction, as well as error propagation during data reduction and model age calculation. The reader is referred to that publication for details and only a summary is provided below.

U-Pb analyses were done at Princeton University following methods described in detail in Barboni *et al.* (53). The zircons were first thermally annealed (48 h at 900 °C), then subjected to a partial dissolution procedure (7, 30, 53) producing two leachates (L1 and L2) and one residue (R) solution, with L1+L2+R corresponding to total digestion. Each zircon was first treated with 100 μL 29 M HF+15 μL 3 M $HNO_3$ for 6 h in a Parr bomb held at 180 °C. The leachate, along with solutions from several rinses of the residual zircon with different acids (particularly HCl, aimed at dissolving fluorides) were collected as solution named L1. Each zircon was then treated a second time under the same conditions (100 μL 29 M HF+15 μL 3 M $HNO_3$ for 6 h at 180 °C). This solution is named L2. The residual zircon was finally dissolved in a Parr bomb using 100 μL 29 M HF+15 μL 3 M $HNO_3$ for 48 h at 210 °C. This solution is named R. Each solution was spiked and equilibrated with the EARTHTIME ($^{202}$Pb-)$^{205}$Pb-$^{233}$U-$^{235}$U spike. Uranium and lead were separated from the rest of the matrix by column chromatography using AG1-X8 resin. The matrix eluted from the U-Pb column was dried down and subsequently redissolved in 0.5 M $HNO_3$ + 0.015 M HF + 1 ppb In. About 30% of the volume was used for analysis of Lu/Hf elemental ratio and trace elements using standard-sample bracketing in an iCAP single collector ICP-MS at Princeton University. The trace element concentrations are given in Datasets S1. The remaining 70% of the solution was dried down and shipped to the Origins Lab of the University of Chicago for Hf purification and isotopic analysis. The Lu/Hf elemental ratios were measured again using standard-sample bracketing by Multi collector ICPMS at the University of Chicago prior to Hf purification. The results agree with iCAP measurements and are compiled in Datasets S1. The Lu/Hf ratios used to correct for $^{176}$Lu-decay in the zircons are those measured at Princeton.

A 2-column procedure was used to purify Hf from all other elements. The first column was filled with 2 mL of TODGA resin (24, 70, 71). The sample was loaded, and the matrix eluted in 20 mL of 12 M $HNO_3$. Titanium was then eluted with 10 mL of 12 M $HNO_3$-1% $H_2O_2$. Iron was then eluted with 10 mL of 3 M $HNO_3$. Hafnium and zirconium were then eluted together in 20 mL of 3 M $HNO_3$-0.3 M HF. The Zr+Hf elution cut was then dried down and loaded in 0.5 mL of 2.5 M HCl on a second column filled with 0.35 mL Ln-Spec (24, 72, 73). The matrix was eluted with 12 mL of 6 M HCl-1% $H_2O_2$. Zirconium was eluted with 22 mL of 6 M HCl-0.06 M HF. Pure Hf was finally eluted in 7 mL of 6 M HCl-0.2 M HF.

Each U-Pb elution was dried down and loaded on a single outgassed zone-refined Re filament with some Si-gel emitter (74) for isotopic analysis using an Isotopx Phoenix Thermal Ionization Mass Spectrometer (TIMS) equipped with ATONA amplifiers at Princeton University (53, 75). Analytical methods were identical to those reported in Barboni *et al*. (53). The calculated U-Pb ages are compiled in Table S1 and Datasets S1.

Hafnium isotopic composition was measured at the Origins Lab of the University of Chicago using a Neptune Multi Collector Inductively Coupled Plasma Mass Spectrometer (MC-ICP-MS) (24) upgraded to Neptune Plus specifications with a Pfeiffer OnTool Booster turbo pump. Purified Hf was introduced into the mass spectrometer in 0.3 M $HNO_3$-0.07 M HF using an Aridus II desolvating nebulizer at an uptake rate of 100 μL/min. Jet sample and X-skimmer cones were used, and the measurements were made in low resolution. Isotopes $^{172}$Yb, $^{174}$Hf, $^{175}$Lu, $^{176}$Hf, $^{177}$Hf, $^{178}$Hf, $^{179}$Hf, $^{180}$Hf, and $^{184}$Hf were measured on Faraday cups connected either to $10^{12}$ Ω amplifiers for $^{172}$Yb and $^{174}$Hf, or $10^{11}$ Ω amplifiers for all other isotopes. The solutions were diluted to 1-10 ppb Hf for analysis. Under optimal conditions, the sensitivity of the instrument was 0.2 V of $^{177}Hf^+$/ppb Hf. Sample measurements were bracketed by analyses of JMC-Hf 475 standard solution diluted to a concentration that matched that of the sample. Bracketing analyses were repeated until the sample solution was nearly empty. *On peak zero* ion intensities were measured on a clean 0.3 M $HNO_3$-0.07 M HF solution. The amounts of Hf analyzed for each zircon and the corresponding spherical-equivalent grain diameter are compiled in Table S1. These are minimum estimates as the zircon grains were plucked out of epoxy mounts and the original grains were larger, having been polished for SIMS analysis. The minimum equivalent diameter is $d=\left(\frac{6m_{\mathrm{Hf}}}{\pi\rho_{\mathrm{zircon}}[\mathrm{Hf}]_{\mathrm{zircon}}}\right)^{1/3}$, where $m_{\mathrm{Hf}}$ is the mass of Hf calculated for original zircons (combination of 30% for trace elements, and 70% for Hf isotope determinations), $\rho_{\mathrm{zircon}}$ is the density of zircon (~4.6 g/cm$^3$), and $[\mathrm{Hf}]_{\mathrm{zircon}}$ is the concentration of Hf in zircon (~1 wt%). Numerically, this takes the form, $d\ [\mu m]=34.63(m_{\mathrm{Hf}}\ [\mathrm{ng}])^{1/3}$.

Isobaric interferences of $^{176}$Yb, $^{176}$Lu, and $^{180}$W were monitored at masses $^{172}$Yb, $^{175}$Lu, and $^{184}$W and were corrected for. They were always negligible. No attempt was made at correcting any contribution of $^{180}$Ta because it is a very low abundance isotope (0.012 %). Hafnium isotopic ratios $^i$Hf/$^{177}$Hf were corrected for instrumental mass fractionation by internal normalization to a fixed $^{179}$Hf/$^{177}$Hf ratio of 0.7325. Hafnium isotopic composition is expressed relative to the value of the JMC-Hf 475, which is used to bracket the sample measurements in a sequence Standard-Sample-Standard,

$$\varepsilon^i\mathrm{Hf}_{\mathrm{sample,j}}=\left[\frac{\left({}^{\mathrm{i}}\mathrm{Hf}/{}^{177}\mathrm{Hf}\right)^*_{\mathrm{sample,j}}}{0.5\left({}^{\mathrm{i}}\mathrm{Hf}/{}^{177}\mathrm{Hf}\right)^*_{\mathrm{standard,j-1}}+0.5\left({}^{\mathrm{i}}\mathrm{Hf}/{}^{177}\mathrm{Hf}\right)^*_{\mathrm{standard,j+1}}}-1\right]\times10^4, \quad (\mathbf{S1})$$

where $({}^{\mathrm{i}}\mathrm{Hf}/{}^{177}\mathrm{Hf})^*$ denotes the internally normalized ratio. When the sample solution was analyzed multiple times during sample-standard bracketing, the average of all bracketed $\varepsilon^i\mathrm{Hf}_{\mathrm{sample,j}}$ values was used. The standard error of this mean was calculated based on the standard deviation of the standards bracketed by themselves. The errors thus calculated were compared by Chen *et al*. (24) to theoretical predictions for the minimum attainable precision imposed by counting statistics and Johnson noise, and the reproducibility achieved was in all cases very close to the theoretical limit. The measured results and propagated errors are compiled in Datasets S1.

## 3. Zircon U-Pb geochronology

The U-Pb ages of 36 zircons were measured in this study in L1, L2, and R. The results are plotted in Fig. S1. The first leachates (L1) were often discordant and not included in this discussion. The second leachates (L2) and residues are mostly near-concordant (within ~0.3% for their $^{207}$Pb/$^{235}$U and $^{206}$Pb/$^{238}$U ages). Some L2 leachates show reversely discordant ($^{206}$Pb/$^{238}$U ages > $^{207}$Pb/$^{235}$U ages), caused by elemental fractionation of U from Pb. As discussed in Barboni *et al*. (53), this is likely an artifact from the partial dissolution process, rather than a problem associated with open system behavior of the zircons. The obtained $^{206}$Pb-$^{207}$Pb ages of L2 are mostly consistent with the residues of the same zircon grain, varying from 3939 to 4338 Ma (Fig. S1).

## 4. Initial ε$^{176}$Hf values

The initial $^{176}$Hf/$^{177}$Hf ratio of a zircon after correction of cosmogenic effects and Lu-decay is calculated relative to the chondritic uniform reservoir (CHUR) in ε-notation ($\varepsilon^{176}\mathrm{Hf}_{\mathrm{zrc}-t,c/CHUR-t}$) as (24),

$$\varepsilon^{176}\text{Hf}_{\text{zrc}-t,c/\text{CHUR}-t} = \left[\frac{\left(\frac{^{176}\text{Hf}}{^{177}\text{Hf}}\right)_{\text{zrc}-p}\left(1-\frac{\alpha_i\varepsilon^i\text{Hf}}{10^4}\right)-\left(\frac{^{176}\text{Lu}}{^{177}\text{Hf}}\right)_{\text{zrc}-p}\left(e^{\lambda_{176}t}-1\right)}{\left(\frac{^{176}\text{Hf}}{^{177}\text{Hf}}\right)_{\text{CHUR-ss}}+\left(\frac{^{176}\text{Lu}}{^{177}\text{Hf}}\right)_{\text{CHUR}-p}\left(e^{\lambda_{176}t_{ss}}-e^{\lambda_{176}t}\right)}-1\right]\times 10^4, \quad \textbf{(S2)}$$

where $\alpha_i$ is the coefficient that relates cosmogenic effects of $\varepsilon^{176}$Hf and $\varepsilon^i$Hf with $i = 178$ and 180 ($\alpha_{178} = +2.35 \pm 0.25$ or $\alpha_{180} = -1.54 \pm 0.11$) (24), $\lambda_{176} = 1.867 \pm 0.008 \times 10^{-11}$ $yr^{-1}$ is the decay constant of $^{176}$Lu calibrated based on isochron analyses of terrestrial rocks with known ages (21) (a more recent estimate based on laboratory decay counting provides a consistent value of 1.864±0.003 $\times 10^{-11}$ $yr^{-1}$ (22)), $t$ is the independently known age from U-Pb dating, $t_{ss} = 4567.30 \pm 0.16$ Myr is the age of the solar system (76), $c$ stands for corrected for cosmogenic effects, $p$ stands for present, $\left(^{176}\text{Hf}/^{177}\text{Hf}\right)_{\text{zrc}-p}$ and $\left(^{176}\text{Lu}/^{177}\text{Hf}\right)_{\text{zrc}-p}$ are ratios measured in the zircon at present, $\left(^{176}\text{Hf}/^{177}\text{Hf}\right)_{\text{CHUR}-ss} = 0.279781 \pm 0.000018$ is the CHUR isotopic composition at the formation of the solar system (73), and $\left(^{176}\text{Lu}/^{177}\text{Hf}\right)_{\text{CHUR}-p} = 0.0338 \pm 0.0001$ is the CHUR Lu/Hf ratio at present (73). The $\left(^{176}\text{Lu}/^{177}\text{Hf}\right)_{\text{CHUR}-p}$ estimate of Izuka *et al.* (73) of $0.0338 \pm 0.0001$ is close to the estimate of Bouvier *et al.* (26) of 0.0336±0.0001. For calculating errors, we follow Chen *et al.* (24) and split the uncertainties originating from zircon analyses from those arising from CHUR by writing,

$$\varepsilon^{176}\text{Hf}_{\text{zrc}-t,c/\text{CHUR}-t} = \varepsilon^{176}\text{Hf}_{\text{zrc}-t,c/\widetilde{\text{CHUR}}-t} - \varepsilon^{176}\text{Hf}_{\text{CHUR}-t/\widetilde{\text{CHUR}}-t} = \left[\frac{\left(\frac{^{176}\text{Hf}}{^{177}\text{Hf}}\right)_{\text{zrc}-p}\left(1-\frac{\alpha_i\varepsilon^i\text{Hf}}{10^4}\right)-\left(\frac{^{176}\text{Lu}}{^{177}\text{Hf}}\right)_{\text{zrc}-p}\left(e^{\lambda_{176}t}-1\right)}{\left(\widetilde{\frac{^{176}\text{Hf}}{^{177}\text{Hf}}}\right)_{\text{CHUR-ss}}+\left(\widetilde{\frac{^{176}\text{Lu}}{^{177}\text{Hf}}}\right)_{\text{CHUR}-p}\left(e^{\tilde{\lambda}_{176}\tilde{t}_{ss}}-e^{\tilde{\lambda}_{176}\tilde{t}}\right)}-1\right]\times 10^4 - \left[\frac{\left(\frac{^{176}\text{Hf}}{^{177}\text{Hf}}\right)_{\text{CHUR-ss}}+\left(\frac{^{176}\text{Lu}}{^{177}\text{Hf}}\right)_{\text{CHUR}-p}\left(e^{\lambda_{176}t_{ss}}-e^{\lambda_{176}t}\right)}{\left(\widetilde{\frac{^{176}\text{Hf}}{^{177}\text{Hf}}}\right)_{\text{CHUR-ss}}+\left(\widetilde{\frac{^{176}\text{Lu}}{^{177}\text{Hf}}}\right)_{\text{CHUR}-p}\left(e^{\tilde{\lambda}_{176}\tilde{t}_{ss}}-e^{\tilde{\lambda}_{176}\tilde{t}}\right)}-1\right]\times 10^4, \quad \textbf{(S3)}$$

where the quantities with the tilde accents take the same values as those without, but the difference is that the former have no error. This allows us to disentangle errors in $\varepsilon^{176}\text{Hf}_{\text{zrc}-t,c/\text{CHUR}-t}$ arising from the CHUR determination from those arising from zircon analysis. This is most useful to evaluate the quality of the measurements independently of CHUR parameters from the literature. It also allows us to discuss model ages of KREEP extraction for zircon populations, treating each data point as independent. While this is not strictly correct because $\lambda_{176}$ and $t$ are present in both $\varepsilon^{176}\text{Hf}_{\text{zrc}-t,c/\text{CHUR}-t}$ and $\varepsilon^{176}\text{Hf}_{\text{zrc}-t,c/\widetilde{\text{CHUR}}-t}$, these terms are known well enough that they represent a very small portion of the overall uncertainty (24). We would not be able to consider zircon Hf isotopic analyses and their uncertainties as independent if the error of CHUR had been incorporated directly in $\varepsilon^{176}\text{Hf}_{\text{zrc}-t,c/\text{CHUR}-t}$ following Eq. S2 (see Chen *et al.* (24) for details). We propagated the uncertainties on all parameters using approximate analytical equations that were verified against the results of Monte-Carlo simulations (24).

## 5. Calculation of model ages

We assume that the lunar magma ocean had a chondritic Lu/Hf ratio until the KREEP reservoir formed with a low Lu/Hf ratio, after which its $^{176}$Hf/$^{177}$Hf evolution diverged from that of CHUR. Zircons crystallized from the KREEP reservoir, either though protracted crystallization or more likely from large scale melting induced by basin-forming impacts (53). The initial $^{176}$Hf/$^{177}$Hf ratios of lunar zircons therefore record the temporal evolution of the $^{176}$Hf/$^{177}$Hf ratio in KREEP. We are interested in dating the time of this departure from CHUR, which we can obtain from zircons in two ways:

(*i*) We can use the initial $^{176}$Hf/$^{177}$Hf ratio and the age of a particular zircon and calculate backwards in time the $^{176}$Hf/$^{177}$Hf ratio of the KREEP reservoir that produced this zircon by assuming a $^{176}$Lu/$^{177}$Hf ratio for KREEP, and by examining when the model KREEP value crossed CHUR. A different model age $t_d$ can be calculated for each zircon, the limitation being that we have to assume $(^{176}\text{Lu}/^{177}\text{Hf})_{\text{KREEP-}p}$ (the present-day elemental ratio of the KREEP reservoir),

$$t_d = \frac{1}{\lambda_{176}}\ln\left[e^{\lambda_{176}t} + \frac{\left(^{176}\text{Hf}/^{177}\text{Hf}\right)_{\text{CHUR}-t}-\left(^{176}\text{Hf}/^{177}\text{Hf}\right)_{\text{zrc}-t,c}}{(^{176}\text{Lu}/^{177}\text{Hf})_{\text{CHUR}-p}-(^{176}\text{Lu}/^{177}\text{Hf})_{\text{KREEP-}p}}\right], \quad \textbf{(S4)}$$

where $(^{176}\mathrm{Hf}/^{177}\mathrm{Hf})_{\mathrm{CHUR}-t}$ and $(^{176}\mathrm{Hf}/^{177}\mathrm{Hf})_{\mathrm{zrc}-t,c}$ are the isotopic ratios of CHUR and the zircon (corrected for cosmogenic effects) at the time of zircon crystallization, respectively, while $(^{176}\mathrm{Lu}/^{177}\mathrm{Hf})_{\mathrm{CHUR}-p} - (^{176}\mathrm{Lu}/^{177}\mathrm{Hf})_{\mathrm{KREEP}\text{-}p}$ are the elemental ratios of CHUR and KREEP at present. Previous studies gave estimates for $(^{176}\mathrm{Lu}/^{177}\mathrm{Hf})_{\mathrm{KREEP}\text{-}p}$ of 0.0164 (25), 0.0154±0.0034 (23), and 0.00153±0.0033 (32). The model ages of zircon calculated in this way are compiled in Datasets S1, assuming a constant $(^{176}\mathrm{Lu}/^{177}\mathrm{Hf})_{\mathrm{KREEP}\text{-}p}$ = 0.00153±0.0033 (32). The uncertainties on the model ages $t_d$ calculated using an approximate analytical equation that was verified against the result of a Monte-Carlo simulation (24) are also compiled in Datasets S1.

(*ii*) If all initial $\varepsilon^{176}$Hf values of zircons plot on a single line, this would be consistent with all of them crystallizing from KREEP with constant $(^{176}\mathrm{Lu}/^{177}\mathrm{Hf})_{\mathrm{KREEP}\text{-}p}$ that separated from CHUR at a precisely defined time. In those circumstances $\varepsilon^{176}\mathrm{Hf}_{\mathrm{zrc}-t,c/\mathrm{CHUR}-t}$ of the zircon population should exhibit the following time-dependence,

$$\varepsilon^{176}\mathrm{Hf}_{\mathrm{zrc}-t,c} - \varepsilon^{176}\mathrm{Hf}_{\mathrm{CHUR}-t} = \frac{10^4\left[(^{176}\mathrm{Lu}/^{177}\mathrm{Hf})_{\mathrm{KREEP}-p} - (^{176}\mathrm{Lu}/^{177}\mathrm{Hf})_{\mathrm{CHUR}-p}\right]\left(e^{\lambda_{176}t_d} - e^{\lambda_{176}t}\right)}{(^{176}\mathrm{Hf}/^{177}\mathrm{Hf})_{\mathrm{CHUR}-ss} + (^{176}\mathrm{Lu}/^{177}\mathrm{Hf})_{\mathrm{CHUR}-p}\left(e^{\lambda_{176}t_{ss}} - e^{\lambda_{176}t}\right)}. \quad \textbf{(S5)}$$

This mathematical relationship is close to linear over the age span of the zircon grains analyzed, and our data indeed can be fit with a linear function, so we can safely use the following approximation at $t_d$,

$$\varepsilon^{176}\mathrm{Hf}_{\mathrm{zrc}-t,c} - \varepsilon^{176}\mathrm{Hf}_{\mathrm{CHUR}-t} \simeq \varepsilon^{176}\mathrm{Hf}_{\mathrm{zrc}-t_d,c} - \varepsilon^{176}\mathrm{Hf}_{\mathrm{CHUR}-t_d} + \left.\frac{\partial\left(\varepsilon^{176}\mathrm{Hf}_{\mathrm{zrc}-t,c} - \varepsilon^{176}\mathrm{Hf}_{\mathrm{CHUR}-t}\right)}{\partial t}\right|_{t=t_d}(t - t_d), \quad \textbf{(S6)}$$

$$\varepsilon^{176}\mathrm{Hf}_{\mathrm{zrc}-t,c} - \varepsilon^{176}\mathrm{Hf}_{\mathrm{CHUR}-t} \simeq \left.\frac{\partial\left(\varepsilon^{176}\mathrm{Hf}_{\mathrm{zrc}-t,c} - \varepsilon^{176}\mathrm{Hf}_{\mathrm{CHUR}-t}\right)}{\partial t}\right|_{t=t_d}(t - t_d), \quad \textbf{(S7)}$$

$$\varepsilon^{176}\mathrm{Hf}_{\mathrm{zrc}-t,c} - \varepsilon^{176}\mathrm{Hf}_{\mathrm{CHUR}-t} \simeq \frac{10^4\left[(^{176}\mathrm{Lu}/^{177}\mathrm{Hf})_{\mathrm{CHUR}-p} - (^{176}\mathrm{Lu}/^{177}\mathrm{Hf})_{\mathrm{KREEP}-p}\right]}{(^{176}\mathrm{Hf}/^{177}\mathrm{Hf})_{\mathrm{CHUR}-ss} + (^{176}\mathrm{Lu}/^{177}\mathrm{Hf})_{\mathrm{CHUR}-p}\left(e^{\lambda_{176}t_{ss}} - e^{\lambda_{176}t_d}\right)}\lambda_{176}(t - t_d), \quad \textbf{(S8)}$$

$$\varepsilon^{176}\mathrm{Hf}_{\mathrm{zrc}-t,c} - \varepsilon^{176}\mathrm{Hf}_{\mathrm{CHUR}-t} \simeq 10^4\left[(^{176}\mathrm{Lu}/^{177}\mathrm{Hf})_{\mathrm{CHUR}-p} - (^{176}\mathrm{Lu}/^{177}\mathrm{Hf})_{\mathrm{KREEP}-p}\right]\left[1 - (^{176}\mathrm{Lu}/^{177}\mathrm{Hf})_{\mathrm{CHUR}-p}\lambda_{176}(t_{ss} - t_d)\right]\lambda_{176}(t - t_d). \quad \textbf{(S9)}$$

The intercept with CHUR at $\varepsilon^{176}\mathrm{Hf}_{\mathrm{zrc}-t,c} - \varepsilon^{176}\mathrm{Hf}_{\mathrm{CHUR}-t} = 0$ gives the age of KREEP $t = t_d$. The slope gives the $(^{176}\mathrm{Lu}/^{177}\mathrm{Hf})_{\mathrm{KREEP}\text{-}p}$ ratio. If we write $\mathcal{s}$ the value of the slope in a diagram $\varepsilon^{176}\mathrm{Hf}_{\mathrm{zrc}-t,c} - \varepsilon^{176}\mathrm{Hf}_{\mathrm{CHUR}-t}$ *vs. t*, we indeed have,

$$(^{176}\mathrm{Lu}/^{177}\mathrm{Hf})_{\mathrm{KREEP}-p} \simeq (^{176}\mathrm{Lu}/^{177}\mathrm{Hf})_{\mathrm{CHUR}-p} - 10^{-4}\mathcal{s}(1/\lambda_{176} + t_{ss} - t_d). \quad \textbf{(S10)}$$

A virtue of this approach is that the $^{176}\mathrm{Lu}/^{177}\mathrm{Hf}$ ratio of KREEP is calculated from the zircon data rather than assumed based on literature data. The caveat is that it only works if KREEP had a relatively simple history, but this assumption can be tested by evaluating the quality of the linear regression, as any departure from a line will indicate either that KREEP was not isolated at a single time, or its $^{176}\mathrm{Lu}/^{177}\mathrm{Hf}$ ratio varied spatially or temporally.

To find the model age and $^{176}\mathrm{Lu}/^{177}\mathrm{Hf}$ of KREEP, we fit $\varepsilon^{176}\mathrm{Hf}_{\mathrm{zrc}-t,c} - \varepsilon^{176}\mathrm{Hf}_{\mathrm{CHUR}-t}$ and $t$ with a line to calculate its intercept with the $x$-axis and its slope. We calculated the MSWD and the data can indeed be fit by a line. Because the CHUR baseline is not independent for all $\varepsilon^{176}\mathrm{Hf}_{\mathrm{zrc}-t,c}$ values, the uncertainty on the model age of KREEP is more accurately calculated by using a Monte Carlo simulation (MCS) (24). Best-fit lines and CHUR baselines were generated following prescribed distributions for the data points and CHUR parameters, and the slopes and intercepts were calculated (24). The intercept ensemble can be described using a binormal distribution that we calculated. We are mostly interested in the marginal probability distribution of $t_d$, which we also calculated and is the basis for our uncertainty estimate. We use the mode of the marginal probability distribution of $t_d$ as best estimate and calculate its 95% confidence interval by taking the 0.025-0.975 interquantile range.

## 6. Comparison between leachate (L2) and residue (R) $^{176}\mathrm{Hf}/^{177}\mathrm{Hf}$ initial values

For zircons for which we have measured both leachate (L2) and residue (R), we are interested in understanding what causes the corrected initial $^{176}\mathrm{Hf}/^{177}\mathrm{Hf}$ values to sometimes differ. The differences between leachate and residue are in all cases small and we can study the cause for the discrepancy by writing the leachate value as a perturbation from the residue value,

$$\varepsilon^{176}\mathrm{Hf}_{\mathrm{L}-t,c/CHUR-t} \simeq \varepsilon^{176}\mathrm{Hf}_{\mathrm{R}-t,c} + \frac{\partial\varepsilon^{176}\mathrm{Hf_R}}{\partial\left({}^{176}\mathrm{Hf}/{}^{177}\mathrm{Hf}\right)_{\mathrm{zrc}-p}}\left({}^{176}\mathrm{Hf}/{}^{177}\mathrm{Hf}_{\mathrm{L}-p} - {}^{176}\mathrm{Hf}/{}^{177}\mathrm{Hf}_{\mathrm{R}-p}\right) + \frac{\partial\varepsilon^{176}\mathrm{Hf_R}}{\partial\varepsilon^{i}\mathrm{Hf}}\left(\varepsilon^{i}\mathrm{Hf}_{L} - \varepsilon^{i}\mathrm{Hf}_{R}\right) + \frac{\partial\varepsilon^{176}\mathrm{Hf_R}}{\partial\left({}^{176}\mathrm{Lu}/{}^{177}\mathrm{Hf}\right)_{\mathrm{zrc}-p}}\left({}^{176}\mathrm{Lu}/{}^{177}\mathrm{Hf}_{L} - {}^{176}\mathrm{Lu}/{}^{177}\mathrm{Hf}_{R}\right) + \frac{\partial\varepsilon^{176}\mathrm{Hf_R}}{\partial e^{\lambda_{176}t}}\left(e^{\lambda_{176}t_L} - e^{\lambda_{176}t_R}\right). \quad \textbf{(S11)}$$

After calculating the partial derivatives and removing the negligible terms, we have,

$$\varepsilon^{176}\mathrm{Hf}_{\mathrm{L}-t,c} - \varepsilon^{176}\mathrm{Hf}_{\mathrm{R}-t,c} \simeq \left(\varepsilon^{176}\mathrm{Hf}_{\mathrm{L}-p} - \varepsilon^{176}\mathrm{Hf}_{\mathrm{R}-p}\right) - \alpha_i\left(\varepsilon^{i}\mathrm{Hf}_{L} - \varepsilon^{i}\mathrm{Hf}_{R}\right) + \frac{10^4\left(1-e^{\lambda_{176}t}\right)}{{}^{176}\mathrm{Hf}/{}^{177}\mathrm{Hf}_{\mathrm{CHUR}-t}}\left({}^{176}\mathrm{Lu}/{}^{177}\mathrm{Hf}_{L} - {}^{176}\mathrm{Lu}/{}^{177}\mathrm{Hf}_{R}\right) + 10^4\,\frac{\left({}^{176}\mathrm{Lu}/{}^{177}\mathrm{Hf}\right)_{\mathrm{CHUR}-p} - \left({}^{176}\mathrm{Lu}/{}^{177}\mathrm{Hf}\right)_{\mathrm{zrc}-p}}{{}^{176}\mathrm{Hf}/{}^{177}\mathrm{Hf}_{\mathrm{CHUR}-t}}\left(e^{\lambda_{176}t_L} - e^{\lambda_{176}t_R}\right). \quad \textbf{(S12)}$$

where $\varepsilon^{176}\mathrm{Hf}_{\mathrm{L}-p}$ and $\varepsilon^{176}\mathrm{Hf}_{\mathrm{R}-p}$ are measured values reported relative to the JMC 475 standard solution. We plot in Fig. S2 the contributions of each term to the overall difference in ${}^{176}\mathrm{Hf}/{}^{177}\mathrm{Hf}$ initial values between leachates and residues.

Zircons can undergo various transformations after formation, including Pb-loss and multi-phase growth (77). Utilizing zircons with these complex histories can lead to data that are difficult to interpret. Additionally, fractionation of the Lu/Hf ratio during chemical abrasion may produce inaccurate initial $\varepsilon^{176}$Hf values (24). To mitigate these issues, we have classified the data into three tiers of reliability. This stratification allows us to filter out disturbed zircons, while maintaining enough original data to yield statistically significant insights into Lu-Hf model ages.

- *Tier 1*: The most reliable indicator of zircon data integrity, reflecting a straightforward history unaffected by laboratory processing, is when L2 and R exhibit similar ages and initial $\varepsilon^{176}$Hf values. We have identified 9 zircons meeting these criteria: 14311_58 z8, 14311_58 z12, 14311_58 z34, 14311_58 z40, 14311_58 z59, 14311_58 z61, 14163_65 z4 (this study), 14311_58 z37 (Barboni *et al.* (28)), and 72275A z1 (Chen *et al.* (24)), comprising 18 U-Pb ages and initial $\varepsilon^{176}$Hf values.
- *Tier 2*: Figure S2 illustrates that in most instances where the initial corrected $\varepsilon^{176}$Hf values differ between L2 and R, the raw $\varepsilon^{176}$Hf values agree. This discrepancy is often due to significant and varying Lu/Hf ratios. It is highly unlikely for different domains within a single zircon to have originated with distinct initial $\varepsilon^{176}$Hf values and Lu/Hf ratios, and then fortuitously converge to similar present-day $\varepsilon^{176}$Hf values after $^{176}$Lu decay. The different initial corrected $\varepsilon^{176}$Hf values are most likely an analytical artifact from fractionation of the Lu/Hf ratio during processing. Chemical abrasion and leaching, intended to remove domains susceptible to Pb-loss, could have caused this fractionation. The residues are most resistant and least affected by Pb-loss. They also contain the majority of Lu and Hf released during chemical abrasion (Table S1). They are therefore likely to retain undisturbed Lu/Hf ratios. In addition to all *Tier 1* measurements, we include in *Tier 2* the residue values with consistent measured (raw) $\varepsilon^{176}$Hf values between leachate and residue. Along with the *Tier 1* zircons, we have identified 9 additional zircons meeting *Tier 2* criteria: 14311_58-z57, 14311_58-z18, 14311_58-z15, 14311_58-z47, 14311_58-z6, 14311_58-z27, 14163_949-z3 (this study), 15405_75-z1 and 14163_65-z7 (Barboni *et al.* (28)). *Tier 2* thus encompasses 27 data points, including 18 from *Tier 1* and 9 additional residue (R) values.
- *Tier 3*: In some instances, insufficient Hf was available to measure L2. All zircons were selected via cathodoluminescence imaging for their probable simple histories (53). Typically, residues, which hold the majority of Lu and Hf, should be less affected by Lu/Hf fractionation during dissolution. Our third approach includes the 27 data points from *Tier 2*, adding the following residues (R): 14311_58-z9, 14311_58-z7, 14311_58-z21, 14311_58-z38, 14311_58-z69a, 14311_58-z69b, 14311_58-z71, 14311_58-z14, 14311_58-z64, 14311_58-z58, 14311_58-z60, 14311_58-z24 (this study), and 14163A-z89 (Chen *et al.*(24)). This corresponds to 13 residue measurements, on top of the 27 counted in *Tier 2*, for a total of 40 data points.

We treat all data as independent in calculation of the number of degrees of freedom and confidence intervals. In reality, because L2 and R are measured on the same zircon, those paired data are correlated to some extent. To assess this effect, we have also performed regressions by calculating the weighted mean and uncertainty for each pair and use these values in the regressions (Fig. S3). The conclusion of this analysis is that L2-R pairing affects minimally the results.

## 7. Comparison with model $^{146}$Sm-$^{142}$Nd and $^{147}$Sm-$^{143}$Nd model ages of mare basalt source formation

A significant argument for a Moon formed relatively late at ~4.33 Ga has been made based on Nd isotopic analyses of KREEP-free mare basalts that provide a $^{146}$Sm-$^{142}$Nd-$^{147}$Sm-$^{143}$Nd model age for the formation of the ultramafic cumulate lithologies that produced the mare basalts (8, 9). Because those data contradict the age that we infer for KREEP, we examine below the assumptions and observations that support such a young age, focusing on the high-quality extensive dataset of Borg *et al.* (8).

We consider a 2-stage model. Time is expressed backwards relative to present, so the solar system formed at $t_{ss}$=4.567 Ga. The Sm/Nd ratio follows a chondritic (CHUR) evolution from the time of solar system formation $t_{ss}$ to the time of lunar magma ocean crystallization and formation of ultramafic cumulate lithologies $t_{LMO}$. The Sm/Nd ratio is then fractionated by crystallization of the lunar magma and formation of mafic cumulate lithologies, which we write $(\mathrm{Sm/Nd})_{\mathrm{UMC}}$. Upon melting of ultramafic cumulate lithologies, mare basalts are formed at time $t_{MB}$ and the Sm/Nd ratio is further fractionated relative to the source. All Sm/Nd ratios are expressed at the present time $t_P$. We are interested in tracking ingrowth of $^{142}$Nd and $^{143}$Nd resulting from decay of $^{146}$Sm ($\lambda_{146} = 6.73 \pm 0.33 \times 10^{-9}$ yr$^{-1}$; $t_{1/2}$=103±5 Myr (8, 78)) and $^{147}$Sm ($\lambda_{147} = 6.539 \times 10^{-12}$ yr$^{-1}$; $t_{1/2}$=106 Gyr). The equations are very similar for the two decay systems, and we use $i$ and $j$ to denote parent and decay product (as in $^{i}$Sm and $^{j}$Nd).

Between $t_{ss}$ and $t_{LMO}$, the Sm/Nd follows a chondritic evolution,

$$\left(\frac{^{j}\mathrm{Nd}}{^{144}\mathrm{Nd}}\right)_{UMC,t_{LMO}} = \left(\frac{^{j}\mathrm{Nd}}{^{144}\mathrm{Nd}}\right)_{\mathrm{CHUR},t_{ss}} + \left(\frac{^{i}\mathrm{Sm}}{^{144}\mathrm{Nd}}\right)_{\mathrm{CHUR},t_{ss}} \left[1 - e^{-\lambda_i (t_{ss}-t_{LMO})}\right], \quad \textbf{(S13)}$$

$$\left(\frac{^{j}\mathrm{Nd}}{^{144}\mathrm{Nd}}\right)_{UMC,t_{LMO}} = \left(\frac{^{j}\mathrm{Nd}}{^{144}\mathrm{Nd}}\right)_{\mathrm{CHUR},t_{ss}} + \left(\frac{^{i}\mathrm{Sm}}{^{147}\mathrm{Sm}}\right)_{CHUR,t_{ss}} \left(\frac{^{147}\mathrm{Sm}}{^{144}\mathrm{Nd}}\right)_{\mathrm{CHUR},t_{ss}} \left[1 - e^{-\lambda_i (t_{ss}-t_{LMO})}\right], \quad \textbf{(S14)}$$

$$\left(\frac{^{j}\mathrm{Nd}}{^{144}\mathrm{Nd}}\right)_{UMC,t_{LMO}} = \left(\frac{^{j}\mathrm{Nd}}{^{144}\mathrm{Nd}}\right)_{\mathrm{CHUR},t_{ss}} + \left(\frac{^{i}\mathrm{Sm}}{^{147}\mathrm{Sm}}\right)_{CHUR,t_{ss}} \left(\frac{^{147}\mathrm{Sm}}{^{144}\mathrm{Nd}}\right)_{\mathrm{CHUR},t_{p}} e^{\lambda_{147} t_{ss}} \left[1 - e^{-\lambda_i (t_{ss}-t_{LMO})}\right], \quad \textbf{(S15)}$$

where $\left(^{j}\mathrm{Nd}/^{144}\mathrm{Nd}\right)_{\mathrm{CHUR},t_{ss}}$ is the initial CHUR ratio at solar system formation, $\left(^{i}\mathrm{Sm}/^{147}\mathrm{Sm}\right)_{t_{ss}}$ is the initial solar system ratio, and $\left(^{147}\mathrm{Sm}/^{144}\mathrm{Nd}\right)_{\mathrm{CHUR},t_p}$ is the present day CHUR ratio ($t_p$=0 Ga in our time reference). To make the comparison easier, we use the same model parameters as Borg *et al.* (8) whenever possible. We use $\left(^{143}\mathrm{Nd}/^{144}\mathrm{Nd}\right)_{\mathrm{CHUR},t_{ss}} = 0.506674$, $\left(^{147}\mathrm{Sm}/^{144}\mathrm{Nd}\right)_{\mathrm{CHUR},t_p} = 0.1967$, and $\left(^{146}\mathrm{Sm}/^{147}\mathrm{Sm}\right)_{t_{ss}} = 0.00164 \pm 0.00009$ [combining $\left(^{146}\mathrm{Sm}/^{144}\mathrm{Sm}\right)_{t_{ss}} = 0.00828 \pm 0.00044$ (78) and $\left(^{147}\mathrm{Sm}/^{144}\mathrm{Sm}\right)_{t_{ss}} = 5.04$] (8). In CHUR, we have,

$$\left(\frac{^{j}\mathrm{Nd}}{^{144}\mathrm{Nd}}\right)_{\mathrm{CHUR},t_{ss}} = \left(\frac{^{j}\mathrm{Nd}}{^{144}\mathrm{Nd}}\right)_{CHUR,t_p} - \left(\frac{^{i}\mathrm{Sm}}{^{147}\mathrm{Sm}}\right)_{CHUR,t_{ss}} \left(\frac{^{147}\mathrm{Sm}}{^{144}\mathrm{Nd}}\right)_{\mathrm{CHUR},t_{p}} e^{\lambda_{147} t_{ss}} \left(1 - e^{-\lambda_i t_{ss}}\right). \quad \textbf{(S16)}$$

We can use the terrestrial standard $^{142}\mathrm{Nd}/^{144}\mathrm{Nd}$ ratio of 1.141837 (79) as a proxy for the present-day CHUR (BSE) ratio to calculate $\left(^{j}\mathrm{Nd}/^{144}\mathrm{Nd}\right)_{\mathrm{CHUR},t_{ss}} = 1.141503$. This is close to previously proposed values of 1.14160±0.00011 (80) and 1.141479±0.000013 (81). Given the possible nucleosynthetic heterogeneity in the Nd isotopic composition of the solar system (82), we prefer to rely on measurements of the terrestrial Nd isotopic composition to estimate $\left(^{j}\mathrm{Nd}/^{144}\mathrm{Nd}\right)_{\mathrm{CHUR},t_{ss}}$ relevant to lunar differentiation. This is a significant source of uncertainty in model ages of lunar magma ocean crystallization based on Nd isotopic analyses of individual mare basalts. As discussed below, using several mare basalt measurements together provides a way to minimize this influence. In the second stage between formation of ultramafic cumulate lithologies and mare basalt formation, we have,

$$\left(\frac{^{j}\mathrm{Nd}}{^{144}\mathrm{Nd}}\right)_{MB,t_{MB}} = \left(\frac{^{j}\mathrm{Nd}}{^{144}\mathrm{Nd}}\right)_{UMC,t_{LMO}} + \left(\frac{^{i}\mathrm{Sm}}{^{144}\mathrm{Nd}}\right)_{UMC,t_{LMO}} \left[1 - e^{-\lambda_i (t_{LMO}-t_{MB})}\right], \quad \textbf{(S17)}$$

$$\left(\frac{^{j}\mathrm{Nd}}{^{144}\mathrm{Nd}}\right)_{MB,t_{MB}} = \left(\frac{^{j}\mathrm{Nd}}{^{144}\mathrm{Nd}}\right)_{UMC,t_{LMO}} + \left(\frac{^{i}\mathrm{Sm}}{^{147}\mathrm{Sm}}\right)_{CHUR,t_{ss}} \left(\frac{^{147}\mathrm{Sm}}{^{144}\mathrm{Nd}}\right)_{UMC,t_{p}} e^{\lambda_{147} t_{SS}} \left[e^{-\lambda_i (t_{ss}-t_{LMO})} - e^{-\lambda_i (t_{SS}-t_{MB})}\right]. \quad \textbf{(S18)}$$

By introducing Eq. S15 into this relationship, we obtain,

$$\left(\frac{{}^{j}\text{Nd}}{{}^{144}\text{Nd}}\right)_{MB,t_{MB}} = \left(\frac{{}^{j}\text{Nd}}{{}^{144}\text{Nd}}\right)_{\text{CHUR},t_{ss}} + \left(\frac{{}^{i}\text{Sm}}{{}^{147}\text{Sm}}\right)_{CHUR,t_{ss}} \left(\frac{{}^{147}\text{Sm}}{{}^{144}\text{Nd}}\right)_{\text{CHUR},t_p} e^{\lambda_{147}t_{SS}}\left[1 - e^{-\lambda_i(t_{ss}-t_{LMO})}\right] + \left(\frac{{}^{i}\text{Sm}}{{}^{147}\text{Sm}}\right)_{CHUR,t_{ss}} \left(\frac{{}^{147}\text{Sm}}{{}^{144}\text{Nd}}\right)_{UMC,t_p} e^{\lambda_{147}t_{SS}}\left[e^{-\lambda_i(t_{ss}-t_{LMO})} - e^{-\lambda_i(t_{SS}-t_{MB})}\right]. \quad \textbf{(S19)}$$

We can write this equation for the two decay systems,

$$\left(\frac{{}^{142}\text{Nd}}{{}^{144}\text{Nd}}\right)_{MB,t_{MB}} = \left(\frac{{}^{142}\text{Nd}}{{}^{144}\text{Nd}}\right)_{\text{CHUR},t_{ss}} + \left(\frac{{}^{146}\text{Sm}}{{}^{147}\text{Sm}}\right)_{CHUR,t_{ss}} \left(\frac{{}^{147}\text{Sm}}{{}^{144}\text{Nd}}\right)_{\text{CHUR},t_p} e^{\lambda_{147}t_{ss}}\left[1 - e^{-\lambda_{146}(t_{ss}-t_{LMO})}\right] + \left(\frac{{}^{146}\text{Sm}}{{}^{147}\text{Sm}}\right)_{CHUR,t_{ss}} \left(\frac{{}^{147}\text{Sm}}{{}^{144}\text{Nd}}\right)_{UMC,t_p} e^{\lambda_{147}t_{SS}}\left[e^{-\lambda_{146}(t_{ss}-t_{LMO})} - e^{-\lambda_{146}(t_{SS}-t_{MB})}\right], \quad \textbf{(S20)}$$

$$\left(\frac{{}^{143}\text{Nd}}{{}^{144}\text{Nd}}\right)_{MB,t_{MB}} = \left(\frac{{}^{143}\text{Nd}}{{}^{144}\text{Nd}}\right)_{\text{CHUR},t_{ss}} + \left(\frac{{}^{147}\text{Sm}}{{}^{144}\text{Nd}}\right)_{\text{CHUR},t_p} \left(e^{\lambda_{147}t_{ss}} - e^{\lambda_{147}t_{LMO}}\right) + \left(\frac{{}^{147}\text{Sm}}{{}^{144}\text{Nd}}\right)_{UMC,t_p} \left(e^{\lambda_{147}t_{LMO}} - e^{\lambda_{147}t_{MB}}\right). \quad \textbf{(S21)}$$

We are interested in relating $^{142}$Nd excesses to the Sm/Nd ratio for the source material of mare basalts. The Sm/Nd ratio can be fractionated during melting and fractional crystallization, but its value in the source can be indirectly inferred from $^{143}$Nd excesses. To do that, we eliminate $\left({}^{147}\text{Sm}/{}^{144}\text{Nd}\right)_{UMC,t_p}$ from the two equations above, and we obtain,

$$\left(\frac{{}^{142}\text{Nd}}{{}^{144}\text{Nd}}\right)_{MB,t_{MB}} = \left(\frac{{}^{142}\text{Nd}}{{}^{144}\text{Nd}}\right)_{\text{CHUR},t_{ss}} + \left(\frac{{}^{146}\text{Sm}}{{}^{147}\text{Sm}}\right)_{CHUR,t_{ss}} \left(\frac{{}^{147}\text{Sm}}{{}^{144}\text{Nd}}\right)_{\text{CHUR},t_p} e^{\lambda_{147}t_{ss}}\left[1 - e^{-\lambda_{146}(t_{ss}-t_{LMO})}\right] + \left(\frac{{}^{146}\text{Sm}}{{}^{147}\text{Sm}}\right)_{CHUR,t_{ss}} \left[\left(\frac{{}^{143}\text{Nd}}{{}^{144}\text{Nd}}\right)_{MB,t_{MB}} - \left(\frac{{}^{143}\text{Nd}}{{}^{144}\text{Nd}}\right)_{\text{CHUR},t_{ss}} - \left(\frac{{}^{147}\text{Sm}}{{}^{144}\text{Nd}}\right)_{\text{CHUR},t_p} \left(e^{\lambda_{147}t_{ss}} - e^{\lambda_{147}t_{LMO}}\right)\right] \left(\frac{e^{\lambda_{146}t_{LMO}} - e^{\lambda_{146}t_{MB}}}{e^{\lambda_{147}t_{LMO}} - e^{\lambda_{147}t_{MB}}}\right) e^{(\lambda_{147}-\lambda_{146})t_{SS}}. \quad \textbf{(S22)}$$

In CHUR, we have,

$$\left(\frac{{}^{j}\text{Nd}}{{}^{144}\text{Nd}}\right)_{CHUR,t_p} = \left(\frac{{}^{j}\text{Nd}}{{}^{144}\text{Nd}}\right)_{\text{CHUR},t_{ss}} + \left(\frac{{}^{i}\text{Sm}}{{}^{147}\text{Sm}}\right)_{CHUR,t_{ss}} \left(\frac{{}^{147}\text{Sm}}{{}^{144}\text{Nd}}\right)_{\text{CHUR},t_p} e^{\lambda_{147}t_{ss}}\left(1 - e^{-\lambda_i t_{ss}}\right). \quad \textbf{(S23)}$$

For a given mare basalt, Sm-Nd isochrons yield initial Nd isotopic compositions $\varepsilon^{142}\text{Nd}_{MB,t_{MB}}$ and $\varepsilon^{143}\text{Nd}_{MB,t_{MB}}$, as well as the age $t_{MB}$. We can thus calculate a model age of LMO differentiation $t_{LMO}$ by solving the above equation for this unknown. Sometimes, only bulk Nd isotopic composition, Sm/Nd, and age are reported. In those cases, the following formula can be used to calculate initial Nd isotopic compositions at the time of mare basalt crystallization,

$$\left(\frac{{}^{j}\text{Nd}}{{}^{144}\text{Nd}}\right)_{MB,t_{MB}} = \left(\frac{{}^{j}\text{Nd}}{{}^{144}\text{Nd}}\right)_{MB,t_p} - \left(\frac{{}^{i}\text{Sm}}{{}^{147}\text{Sm}}\right)_{t_{ss}} \left(\frac{{}^{147}\text{Sm}}{{}^{144}\text{Nd}}\right)_{MB,t_p} e^{\lambda_{147}t_{ss}}\left[e^{-\lambda_i(t_{ss}-t_{MB})} - e^{-\lambda_i t_{ss}}\right]. \quad \textbf{(S24)}$$

Several studies have reported mare basalt Nd isotopic compositions (8, 83-85). The data considered here are from Borg *et al*. (8), as these were all measured using the same protocol and show less dispersion than other data sets (9). Borg *et al*. (8) reported a model age for the source of non-KREEP mare basalts of 4.336±0.030 Myr, which was subsequently revised to 4.331±0.014 Myr (9). We plot in Fig. S5 the predicted $^{142}$Nd/$^{144}$Nd initial ratios of mare basalts (at the time of eruption/solidification) with the measured $^{142}$Nd/$^{144}$Nd initial ratios. The predicted $^{142}$Nd/$^{144}$Nd initial ratios are calculated based on $^{143}$Nd/$^{144}$Nd initial ratios measured in the same samples. We leave the CHUR (bulk silicate Moon) $^{142}$Nd/$^{144}$Nd initial ratio and time of LMO differentiation as adjustable parameters. As expected, we confirm the finding of Borg and Carlson (9) and references therein that the Nd isotopic data are best explained by an age of 4.33 Ga. All data cannot however be explained by a single age. We also plot the predicted $^{142}$Nd/$^{144}$Nd mare basalt initial values for an age of LMO differentiation of 4.44 Ga. The match between observed and predicted values is not as good as with a model age of 4.33 Ga.

To summarize, we agree that the model age for the source of mare basalts that best fits the data is 4.33 Ga, but many data are unexplained indicating incomplete resetting. As discussed in the main text, this age could reflect incipient melting or density-driven mixing among the cumulate lithologies that would later form the mare basalts.

## 8. Zircon saturation in KREEP magma

The zircons analyzed are from regolith lithologies and lack petrologic context. Some of them could have crystallized from KREEP-rich basalts erupted at the lunar surface, while others could have formed by cooling of the KREEP magma reservoir (86). In Fig. S6, we calculate the degree of crystallization and temperature at which zircons would crystallize from a KREEP magma (27) using the MELTS model (87) and a zircon saturation model (88). Under this simplistic scenario, it would take 88% crystallization for the magma to reach saturation. This means that the lunar magma ocean could have reached 99.9% crystallization before lunar zircons started crystallizing (99% to make KREEP, and 90% crystallization of the 1% remaining KREEP magma to saturate zircons). LMO crystallization is not the only event that could have formed zircons. Crystallization of erupted lavas rich in KREEP and secondary melting by impacts could have played significant roles as well.

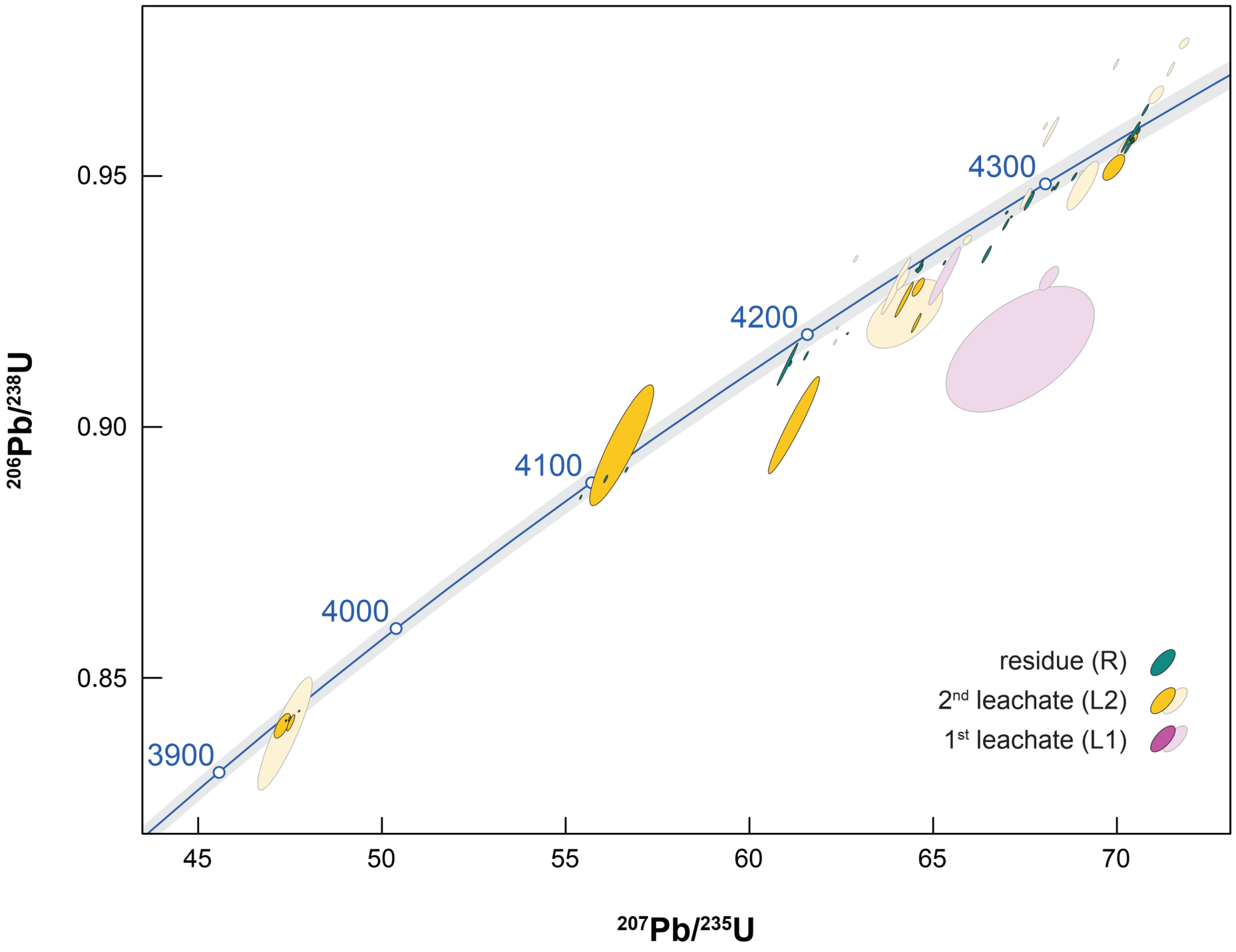

**Fig. S1.** Wetherill concordia diagram of all U–Pb ID-TIMS zircon dates (53) used to determine Hf model ages in this study, separated into residue (R), second leachate (L2) and first leachate (L1) aliquots. Leachates and residues classified as Tier 1, 2, and 3 are depicted in solid colors, whereas other leachates are shown in transparent colors.

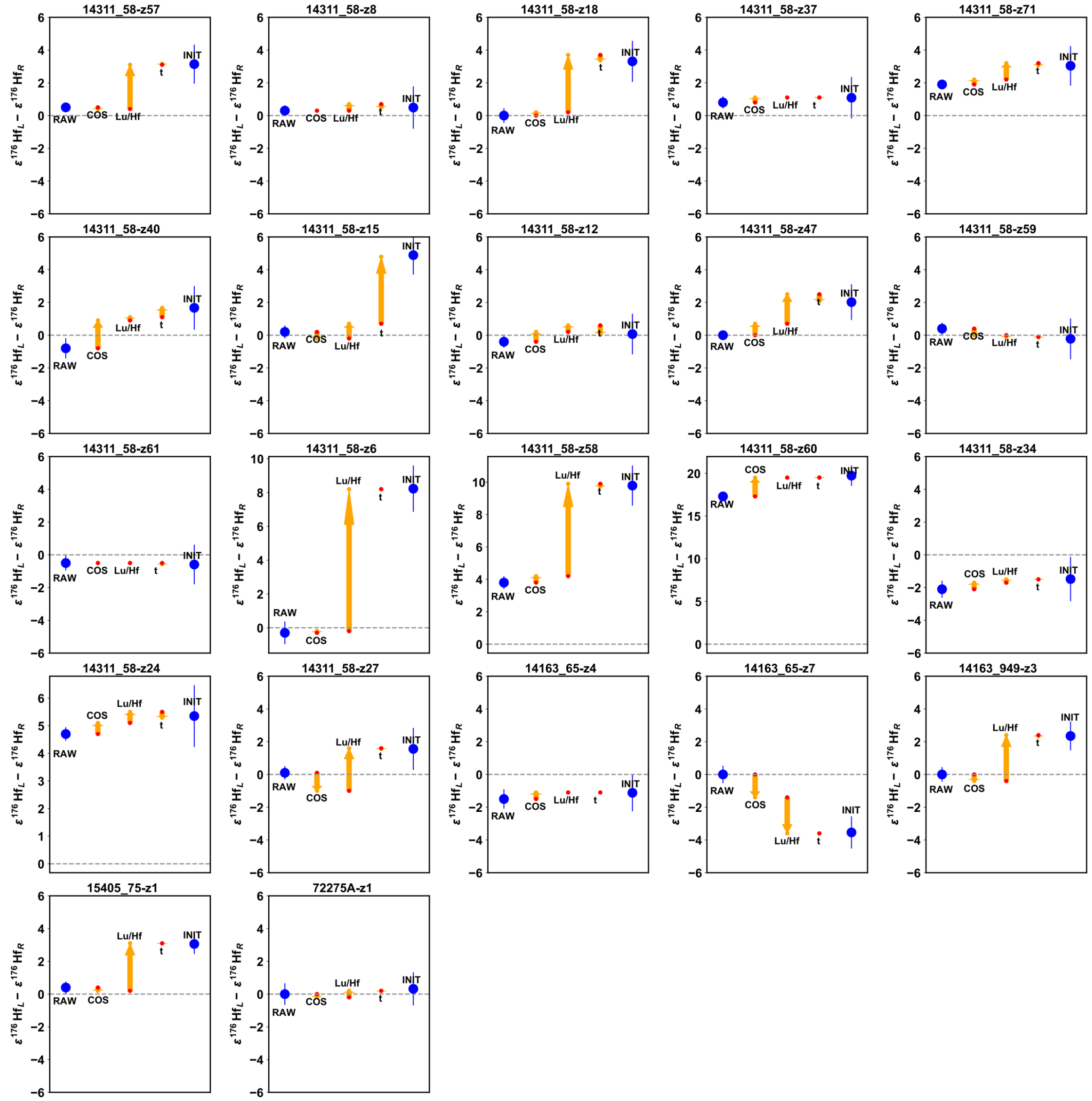

**Fig. S2.** Analysis of the origin of the difference in initial $\varepsilon^{176}$Hf values between leachate and residue based on Eq. S12. Each panel corresponds to a different zircon for which L2 and R data are available (the name of each zircon is indicated at the top of each panel). The blue filled dot on the left of each panel is the measured (RAW) isotopic difference between leachate and residue $\varepsilon^{176}\mathrm{Hf}_{\mathrm{L-p}} - \varepsilon^{176}\mathrm{Hf}_{\mathrm{R-p}}$. Each yellow arrow indicates how differences in the correction of cosmogenic effects (COS), parent/product ratio ($^{176}$Lu/$^{177}$Hf), time (t) influence the overall difference in initial $\varepsilon^{176}$Hf values after correction of cosmogenic effects and in situ $^{176}$Lu-decay. The difference in initial corrected $\varepsilon^{176}$Hf values between leachate and residue is shown as a blue filled dot on the right of each panel (INIT).

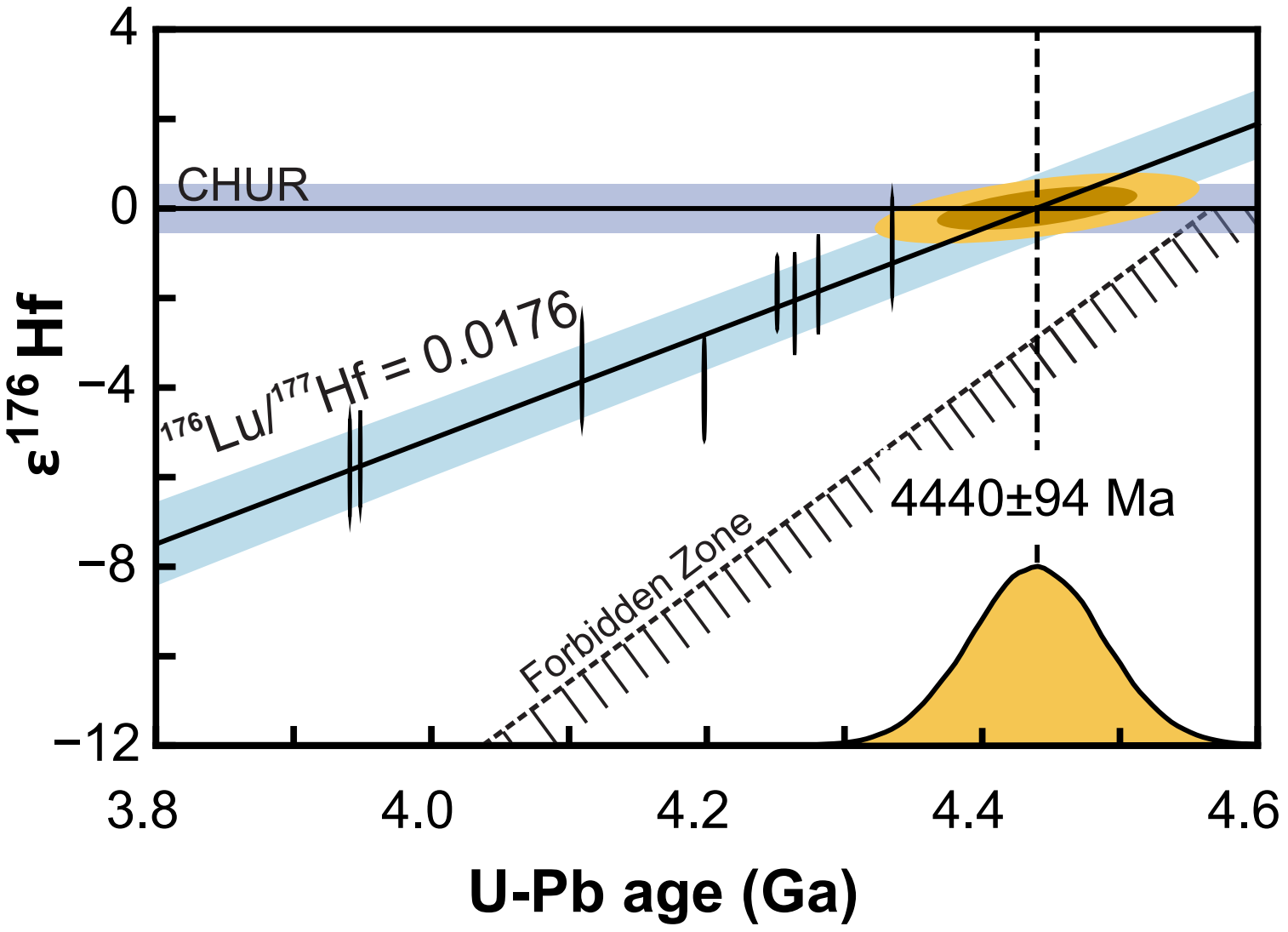

**Fig. S3.** Weighted mean initial $\varepsilon^{176}$Hf values (cosmogenic effects were corrected using $\varepsilon^{178}$Hf) and U/Pb ages of leachates L2 and residues for Tier 1 lunar zircons.

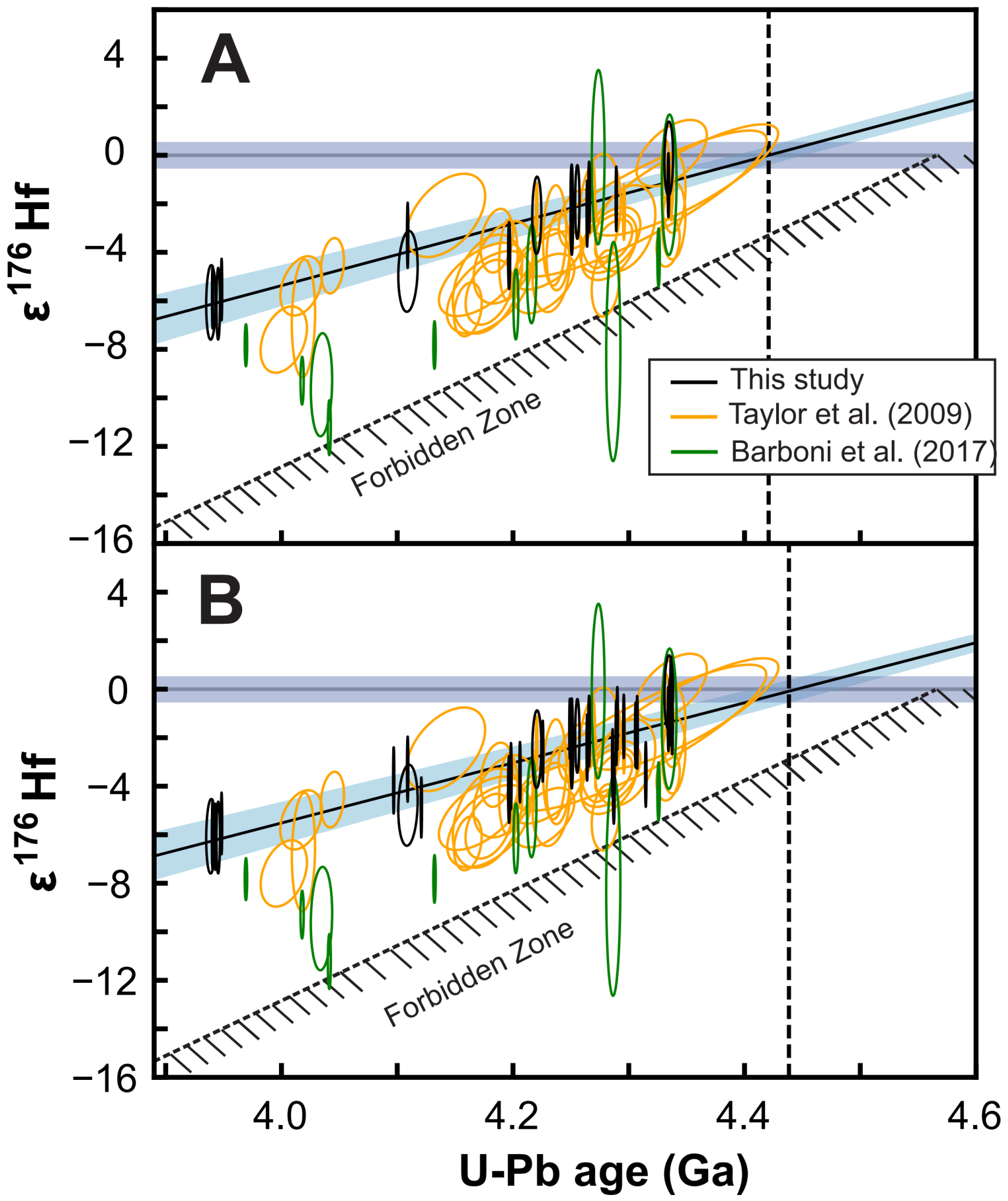

**Fig. S4.** Comparison between present analyses of initial $\varepsilon^{176}$Hf values (cosmogenic effects were corrected using $\varepsilon^{178}$Hf) and U-Pb ages of lunar zircons (Table S1) and previous studies (7, 25). The top panel (A) shows Tier 1 data, while the bottom panel (B) shows Tier 3 data. As illustrated, the new data are significantly more precise and show significantly less scatter than previous data.

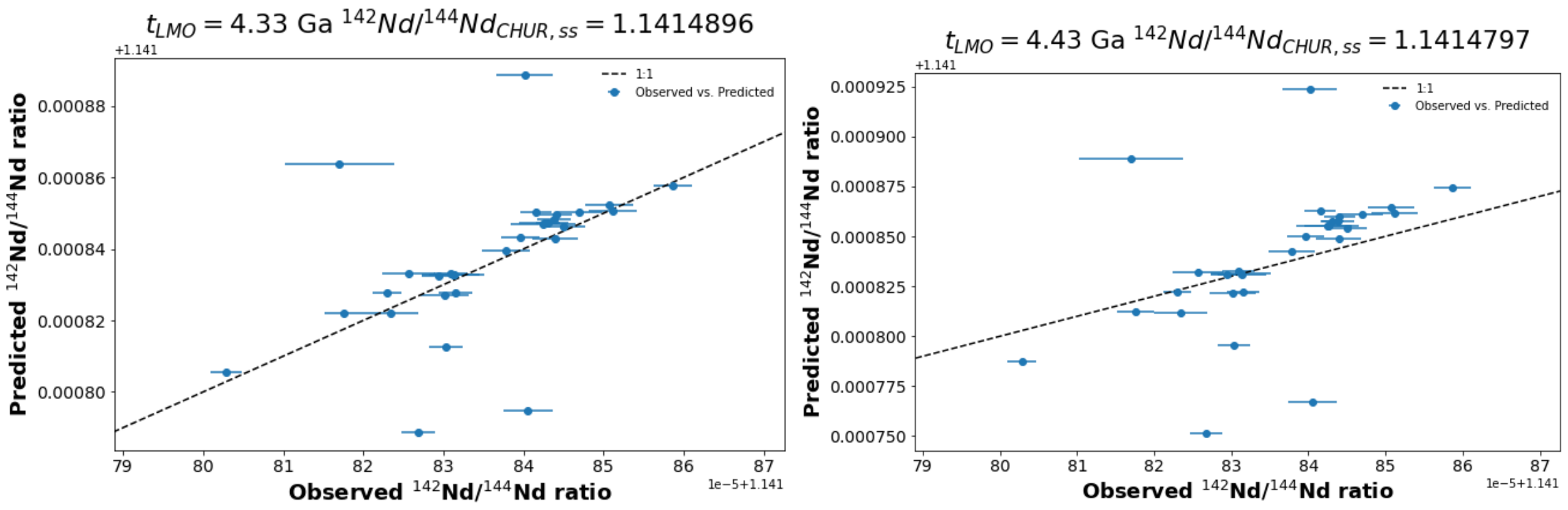

**Fig. S5**. Comparison between predicted and observed $^{142}$Nd/$^{144}$Nd initial ratios of mare basalts using a 2-stage model (Eq. S22) for different model ages of LMO crystallization (4.33 Ga on the left (9); 4.43 Ga on the right; this study). In each case, the CHUR (bulk silicate Moon) $^{142}$Nd/$^{144}$Nd initial ratio at the time of solar system formation was adjusted to fit the data. The predicted mare basalt $^{142}$Nd/$^{144}$Nd initial ratios were calculated using measured $^{143}$Nd/$^{144}$Nd initial ratios. Data from Borg *et al*. (8).

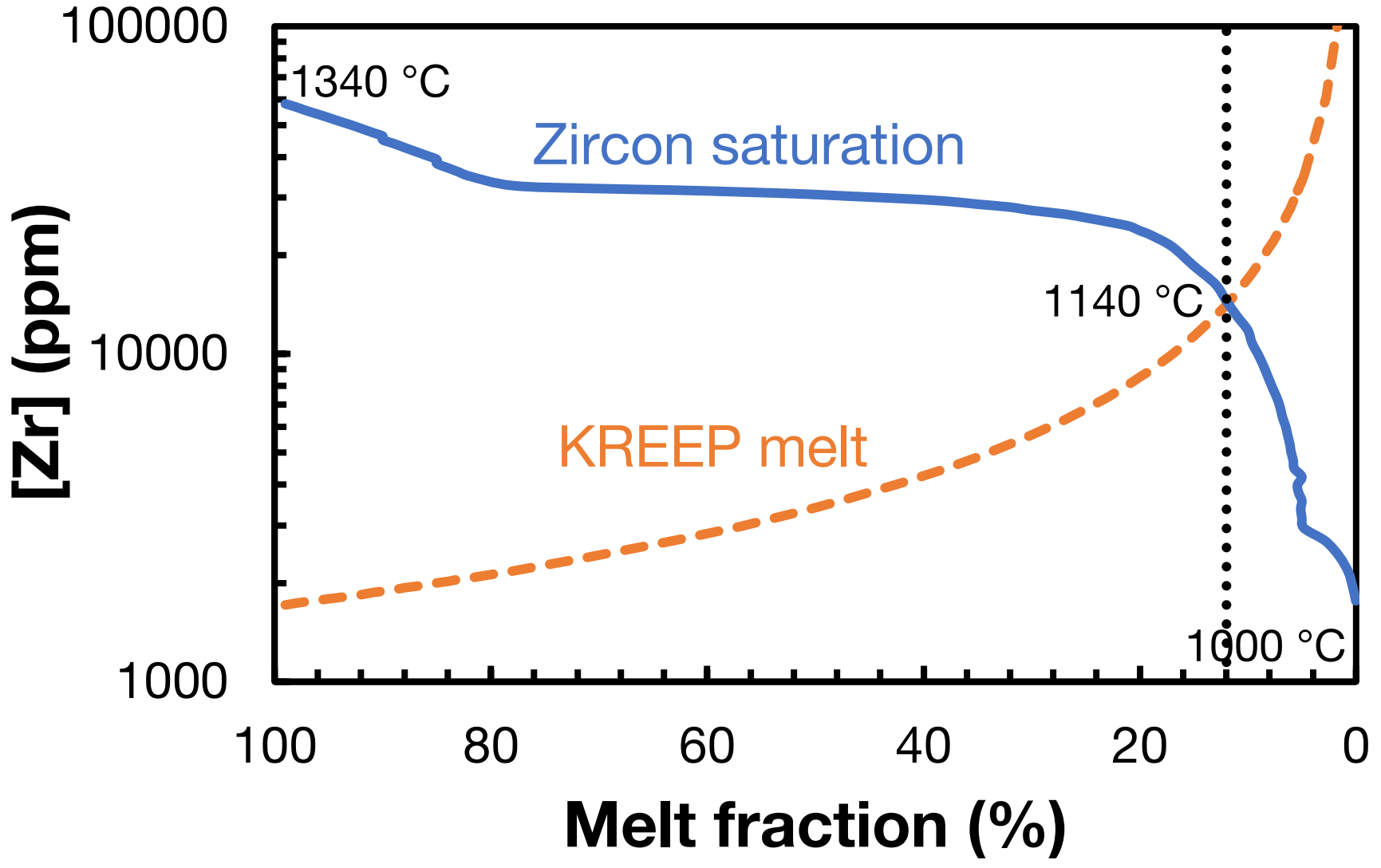

**Fig. S6.** Calculation of zircon saturation in a KREEP magma during equilibrium crystallization. The blue solid curve is the zircon saturation curve (88) for equilibrium crystallization of anhydrous bulk KREEP composition (27) at 5 kbar based on the MELTS model (87). The KREEP melt curve assumes complete incompatible behavior of Zr, starting with the Zr concentration of KREEP from Warren and Wasson (27). As shown, crystallizing KREEP liquid will not saturate zircon until the melt is 88% crystallized (melt fraction=12 %) at 1140 °C.

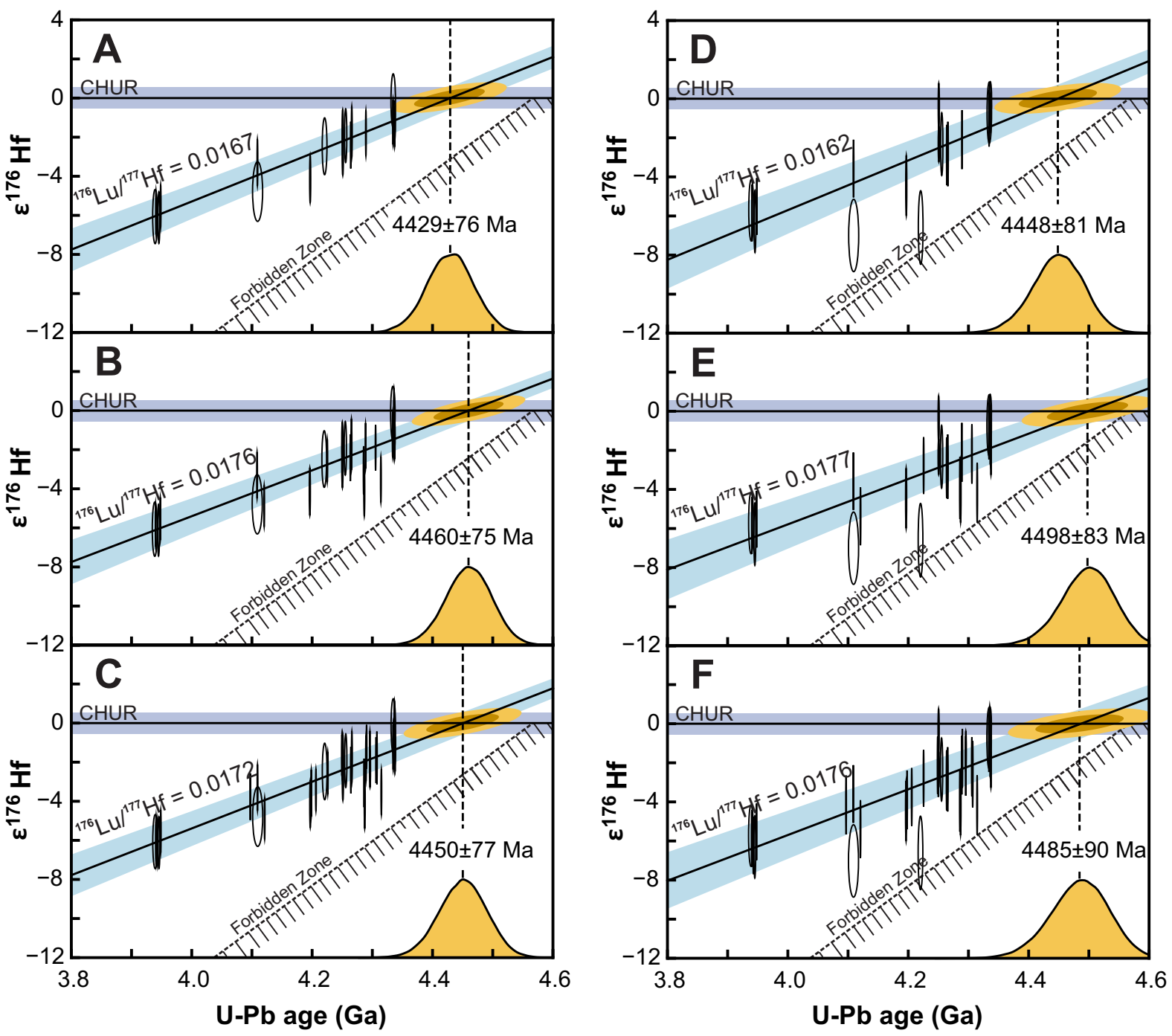

**Fig. S7.** Comparison between Tier 1 (A, D), Tier 2 (B, E) and Tier 3 (C, F) data sets correcting cosmogenic effects either using $\varepsilon^{178}$Hf (A, B, C) or $\varepsilon^{180}$Hf (D, E, F). Details on the latter correction are provided in the text and Chen *et al*. (24). As shown, all KREEP model ages and $^{176}$Lu/$^{177}$Hf ratios inferred from the data agree. Correction using $\varepsilon^{180}$Hf introduces some scatter, the origin of which is uncertain.

**Table S1.** Summary of U-Pb ages and Hf isotopic compositions of lunar zircons

| Sample | Tier # | Hf (ng) | Zircon diameter (μm) | $\varepsilon^{176}Hf_{CHUR}$ corrected with $^{178}Hf$ | 2σ* | $\varepsilon^{176}Hf_{CHUR}$ corrected with $^{180}Hf$ | 2σ* | Lu/Hf (ppm/ppm) | 2σ | Pb-Pb Ages (Ma)[a] | 2σ | Ref (Ages) | Ref ($\varepsilon^{176}Hf$) |
|---|---|---|---|---|---|---|---|---|---|---|---|---|---|
| 14311_58_z8_R | 1,2,3 | 17.4 | 90 | -2.41 | 1.36 | -2.90 | 1.7 | 0.0068 | 0.0002 | 4264.04 | 0.50 | 1 | 1 |
| 14311_58_z8_L2 | 1,2,3 | 4.9 | 59 | -1.92 | 1.51 | -2.34 | 1.7 | 0.0059 | 0.0002 | 4255.58 | 1.86 | 1 | 1 |
| 14311_58_z37_R | 1,2,3 | 11.7 | 79 | -1.23 | 1.31 | -1.03 | 1.6 | 0.0055 | 0.0001 | 4334.24 | 0.56 | 2 | 2 |
| 14311_58_z37_L2 | 1,2,3 | 3.9 | 55 | -0.14 | 1.53 | -0.73 | 1.7 | 0.0054 | 0.0002 | 4334.55 | 3.74 | 2 | 2 |
| 14311_58_z40_R | 1,2,3 | 12.6 | 80 | -4.14 | 1.38 | -4.46 | 1.7 | 0.0056 | 0.0002 | 4196.64 | 0.88 | 1 | 1 |
| 14311_58_z40_L2 | 1,2,3 | 1.2 | 37 | -2.47 | 1.59 | -6.61 | 2.0 | 0.0051 | 0.0002 | 4220.57 | 3.91 | 1 | 1 |
| 14311_58_z12_R | 1,2,3 | 11.9 | 79 | -1.80 | 1.33 | -2.11 | 1.6 | 0.0066 | 0.0003 | 4289.21 | 0.54 | 1 | 1 |
| 14311_58_z12_L2 | 1,2,3 | 4.2 | 56 | -1.73 | 1.46 | -2.84 | 1.8 | 0.0056 | 0.0002 | 4265.73 | 0.71 | 1 | 1 |
| 14311_58_z59_R | 1,2,3 | 164.6 | 190 | -5.85 | 1.29 | -5.86 | 1.6 | 0.0069 | 0.0003 | 3940.48 | 0.50 | 1 | 1 |
| 14311_58_z59_L2 | 1,2,3 | 4.0 | 55 | -6.07 | 1.53 | -5.72 | 1.7 | 0.0072 | 0.0003 | 3939.08 | 4.10 | 1 | 1 |
| 14311_58_z61_R | 1,2,3 | 210.5 | 206 | -5.53 | 1.27 | -5.53 | 1.6 | 0.0070 | 0.0001 | 3948.29 | 0.50 | 1 | 1 |
| 14311_58_z61_L2 | 1,2,3 | 5.5 | 61 | -6.12 | 1.47 | -6.30 | 1.7 | 0.0070 | 0.0002 | 3945.34 | 1.53 | 1 | 1 |
| 14311_58_z34_R | 1,2,3 | 20.0 | 94 | -3.30 | 1.34 | -3.59 | 1.6 | 0.0081 | 0.0002 | 4109.00 | 0.57 | 1 | 1 |
| 14311_58_z34_L2 | 1,2,3 | 1.4 | 38 | -4.78 | 1.67 | -7.01 | 2.0 | 0.0076 | 0.0007 | 4109.28 | 8.35 | 1 | 1 |
| 14163_65_z4_R | 1,2,3 | 14.1 | 84 | -1.52 | 1.13 | -0.40 | 1.3 | 0.0074 | 0.0002 | 4250.67 | 0.94 | 1 | 1 |
| 14163_65_z4_L2 | 1,2,3 | 2.0 | 44 | -2.64 | 1.47 | -2.34 | 1.1 | 0.0074 | 0.0004 | 4250.79 | 1.07 | 1 | 1 |
| 72275A_z1_R | 1,2,3 | 19.6 | 93 | -1.54 | 0.33 | -0.72 | 0.36 | 0.0070 | 0.0002 | 4336.84 | 0.50 | 3 | 3 |
| 72275A_z1_L2 | 1,2,3 | 16.2 | 88 | -1.22 | 0.38 | -0.07 | 0.31 | 0.0061 | 0.0007 | 4336.22 | 2.12 | 3 | 3 |
| 14311_58_z57_R | 2,3 | 13.8 | 83 | -3.52 | 1.33 | -4.16 | 1.70 | 0.0141 | 0.0006 | 4314.69 | 0.57 | 2 | 1 |
| 14311_58_z57_L2 | | 19.6 | 93 | -0.38 | 1.37 | -0.93 | 1.70 | 0.0078 | 0.0004 | 4317.55 | 0.55 | 2 | 1 |
| 14311_58_z15_R | 2,3 | 57.3 | 133 | -4.88 | 1.24 | -5.38 | 1.60 | 0.0056 | 0.0003 | 4120.84 | 0.52 | 1 | 1 |
| 14311_58_z15_L2 | | 7.3 | 67 | 0.01 | 1.47 | 0.28 | 1.7 | 0.0033 | 0.0002 | 4293.17 | 1.16 | 1 | 1 |
| 14311_58_z47_R | 2,3 | 58.0 | 134 | -2.56 | 1.25 | -2.79 | 1.6 | 0.0093 | 0.0003 | 4225.71 | 0.49 | 1 | 1 |
| 14311_58_z47_L2 | | 10.9 | 77 | -0.54 | 1.27 | -1.59 | 1.7 | 0.0049 | 0.0001 | 4207.26 | 0.61 | 1 | 1 |
| 14311_58_z6_R | 2,3 | 5.0 | 59 | -4.09 | 1.45 | -4.12 | 1.7 | 0.0379 | 0.0006 | 4287.09 | 0.77 | 1 | 1 |
| 14311_58_z6_L2 | | 2.2 | 45 | 4.14 | 1.57 | 2.93 | 2.0 | 0.0179 | 0.0008 | 4287.33 | 0.85 | 1 | 1 |
| 14311_58_z27_R | 2,3 | 26.7 | 104 | -1.89 | 1.34 | -2.19 | 1.6 | 0.0147 | 0.0007 | 4305.90 | 0.49 | 2 | 1 |
| 14311_58_z27_L2 | | 5.4 | 61 | -0.33 | 1.52 | 1.36 | 1.8 | 0.0085 | 0.0004 | 4304.50 | 0.57 | 2 | 1 |
| 14311_58_z18_R | 2,3 | 25.3 | 102 | -3.06 | 1.40 | -3.93 | 1.7 | 0.0225 | 0.0007 | 4286.02 | 0.50 | 1 | 1 |
| 14311_58_z18_L2 | | 6.8 | 65 | 0.25 | 1.40 | -0.73 | 1.8 | 0.0142 | 0.0005 | 4270.90 | 1.68 | 1 | 1 |
| 14163_65_z7_R | 2,3 | 4.1 | 56 | 0.15 | 1.18 | -1.64 | 1.3 | 0.0144 | 0.0005 | 4336.46 | 0.59 | 2 | 2 |
| 14163_65_z7_L2 | | 8.5 | 71 | -3.39 | 1.16 | -2.5 | 1.1 | 0.0196 | 0.0004 | 4336.58 | 1.70 | 2 | 2 |
| 14163_949_z3_R | 2,3 | 20.8 | 95 | -1.66 | 1.03 | -2.34 | 1.0 | 0.0145 | 0.0003 | 4337.58 | 0.51 | 1 | 1 |
| 14163_949_z3_L2 | | 5.9 | 63 | 0.69 | 1.15 | 0.87 | 1.0 | 0.008 | 0.0005 | 4334.88 | 2.30 | 1 | 1 |
| 15405_75_z1_R | 2,3 | 14.9 | 85 | -1.09 | 0.99 | -1.15 | 1.0 | 0.0162 | 0.0006 | 4337.32 | 0.51 | 2 | 2 |
| 15405_75_z1_L2 | | 4.6 | 57 | 1.97 | 0.80 | 2.31 | 1.0 | 0.0094 | 0.0004 | 4336.91 | 0.66 | 2 | 2 |
| 14311_58_z9_R | 3 | 10.3 | 75 | -0.95 | 1.36 | -0.55 | 1.6 | 0.0088 | 0.0003 | 4335.49 | 0.75 | 2 | 2 |
| 14311_58_z7_R | 3 | 6.1 | 63 | -1.78 | 1.49 | -2.57 | 1.8 | 0.0102 | 0.0004 | 4307.26 | 0.53 | 2 | 1 |
| 14311_58_z21_R | 3 | 3.5 | 52 | -2.53 | 1.55 | -3.41 | 1.8 | 0.0206 | 0.0011 | 4334.94 | 0.54 | 2 | 2 |
| 14311_58_z38_R | 3 | 12.6 | 81 | -1.53 | 1.31 | -2.15 | 1.6 | 0.01 | 0.0004 | 4295.71 | 0.53 | 1 | 1 |
| 14311_58_z43_L2 | | 1.8 | 42 | -1.19 | 1.71 | -0.24 | 1.8 | 0.0093 | 0.0004 | 4322.34 | 4.02 | 2 | 1 |
| 14311_58_z69b_R | 3 | 29.2 | 107 | -3.59 | 1.36 | -3.89 | 1.6 | 0.0055 | 0.0003 | 4198.27 | 0.52 | 1 | 1 |
| 14311_58_z69a_R | 3 | 45.2 | 123 | -3.42 | 1.24 | -3.74 | 1.6 | 0.0055 | 0.0003 | 4206.03 | 0.53 | 1 | 1 |
| 14311_58_z71_R | 3 | 67.0 | 141 | -1.64 | 1.23 | -2.13 | 1.6 | 0.0095 | 0.0003 | 4249.09 | 0.49 | 1 | 1 |
| 14311_58_z71_L2 | | 4.6 | 58 | 1.4 | 1.51 | 0.59 | 1.8 | 0.0071 | 0.0006 | 4243.67 | 1.79 | 1 | 1 |
| 14311_58_z14_R | 3 | 25.1 | 101 | -3.76 | 1.36 | -4.15 | 1.6 | 0.0094 | 0.0003 | 4096.95 | 0.51 | 1 | 1 |
| 14311_58_z64_R | 3 | 146.4 | 183 | -6.16 | 1.27 | -5.89 | 1.5 | 0.007 | 0.0002 | 3940.42 | 0.49 | 1 | 1 |
| 14311_58_z58_R | 3 | 56.8 | 133 | -1.22 | 1.30 | -1.77 | 1.6 | 0.0303 | 0.0005 | 4290.14 | 0.49 | 1 | 1 |
| 14311_58_z58_L2 | | 7.9 | 69 | 8.57 | 1.48 | 8.71 | 1.8 | 0.0169 | 0.0005 | 4282.34 | 0.53 | 1 | 1 |
| 14311_58_z60_R | 3 | 28.1 | 105 | -6.09 | 1.36 | -6.23 | 1.6 | 0.0075 | 0.0002 | 3942.92 | 0.54 | 1 | 1 |
| 14311_58_z60_L2 | | 1.1 | 35 | 13.64 | 1.36 | 14.08 | 1.4 | 0.0075 | 0.0003 | 3944.21 | 8.83 | 1 | 1 |
| 14311_58_z24_R | 3 | 39.8 | 118 | -1.72 | 1.33 | -2.21 | 1.7 | 0.0069 | 0.0006 | 4250.18 | 0.55 | 1 | 1 |
| 14311_58_z24_L2 | | 3.5 | 52 | 3.63 | 1.26 | 2.87 | 1.6 | 0.0059 | 0.0002 | 4241.43 | 1.70 | 1 | 1 |
| 14163_65_z3_L2 | | 3.7 | 53 | -1.08 | 0.94 | -1.1 | 1.1 | 0.0121 | 0.0006 | 4336.53 | 2.51 | 2 | 2 |
| 14163A_z9_L1 | | 9.5 | 73 | -1.50 | 0.40 | -1.24 | 0.29 | 0.0138 | 0.0005 | 4268.33 | 2.44 | 3 | 3 |
| 14163A_z26_L1 | | 9.0 | 72 | 0.07 | 0.47 | 0.89 | 0.58 | 0.0046 | 0.0002 | 4337.07 | 30.35 | 3 | 3 |
| 14163A_z26_L2 | | 10.1 | 75 | -2.64 | 0.37 | -1.76 | 0.44 | 0.0044 | 0.0001 | 4255.67 | 16.19 | 3 | 3 |
| 14163A_z89_R | 3 | 9.2 | 73 | -2.34 | 0.33 | -1.42 | 0.37 | 0.0058 | 0.0002 | 4295.85 | 0.83 | 3 | 3 |
| 14321A_z3_L1 | | 71.4 | 144 | -2.17 | 0.22 | -1.87 | 0.33 | 0.0105 | 0.0017 | 4220.48 | 0.60 | 3 | 3 |
| 14321A_z3_L2 | | 40.3 | 119 | -3.80 | 0.41 | -3.35 | 0.28 | 0.0146 | 0.0006 | 4217.48 | 0.55 | 3 | 3 |
| 72275A_z1_L1 | | 22.3 | 98 | -0.60 | 0.35 | -0.55 | 0.29 | 0.0064 | 0.0003 | 4331.63 | 3.32 | 3 | 3 |

*Errors include uncertainties from both CHUR and zircon measurements. See the Excel spreadsheet in Datasets S1 for details on the data reduction.

[a] $^{207}Pb/^{206}Pb$ ages corrected for initial Th/U disequilibrium using radiogenic $^{208}Pb$ and Th/U[magma] = 3.50000.

References: 1. This study; 2. Barboni *et al*. (53); 3. Chen *et al*. (24)

## SI References

1. D. J. Taylor, K. D. McKeegan, T. M. Harrison, Lu–Hf zircon evidence for rapid lunar differentiation. *Earth and Planetary Science Letters* **279**, 157-164 (2009).
2. X. Chen *et al.*, Methodologies for $^{176}$Lu–$^{176}$Hf analysis of zircon grains from the Moon and beyond. *ACS Earth and Space Chemistry* 10.1021/acsearthspacechem.3c00093 (2023).
3. M. Barboni *et al.*, High precision U-Pb zircon dating pinpoints a major magmatic event on the Moon at 4.337 Ga. *Science Advances* **10**, eadn9871 (2024).
4. C. Meyer, I. S. Williams, W. Compston, Uranium-lead ages for lunar zircons: Evidence for a prolonged period of granophyre formation from 4.32 to 3.88 Ga. *Meteoritics & Planetary Science* **31**, 370-387 (1996).
5. C. Meyer, Lunar sample compendium. (2005).
6. L. A. Haskin, R. L. Korotev, J. J. Gillis, B. L. Jolliff (2002) Stratigraphies of Apollo and Luna highland landing sites and provenances of materials from the perspective of basin impact ejecta modeling. in *Lunar and Planetary Science Conference*, p 1364.
7. R. Morrison, V. Oberbeck (1975) Geomorphology of crater and basin deposits-Emplacement of the Fra Mauro formation. in *Lunar Science Conference, 6th, Houston, Tex., March 17-21, 1975, Proceedings. Volume 3.(A78-46741 21-91) New York, Pergamon Press, Inc., 1975, p. 2503-2530*.
8. F. J. Stadermann, E. Heusser, E. K. Jessberger, S. Lingner, D. Stöffler, The case for a younger Imbrium basin: New $^{40}$Ar-$^{39}$Ar ages of Apollo 14 rocks. *Geochimica et Cosmochimica Acta* **55**, 2339-2349 (1991).
9. R. Drozd, C. Hohenberg, C. Morgan, C. Ralston, Cosmic-ray exposure history at the Apollo 16 and other lunar sites: lunar surface dynamics. *Geochimica et Cosmochimica Acta* **38**, 1625-1642 (1974).
10. R. E. Merle *et al.*, Origin and transportation history of lunar breccia 14311. *Meteoritics & Planetary Science* **52**, 842-858 (2017).
11. M. Hopkins, S. J. Mojzsis, A protracted timeline for lunar bombardment from mineral chemistry, Ti thermometry and U–Pb geochronology of Apollo 14 melt breccia zircons. *Contributions to Mineralogy and Petrology* **169**, 1-18 (2015).
12. J. M. Devine, D. S. McKay, J. J. Papike, Lunar regolith: Petrology of the $<10$ µm fraction. *Journal of Geophysical Research: Solid Earth* **87**, A260-A268 (1982).
13. T. Labotka, M. Kempa, C. White, J. Papike, J. Laul (1980) The lunar regolith-Comparative petrology of the Apollo sites. *In: Lunar and Planetary Science Conference, 11th, Houston, TX, March 17-21, 1980, Proceedings. Volume 2.(A82-22296 09-91) New York, Pergamon Press, 1980, p. 1285-1305.*
14. C. A. Crow, K. D. McKeegan, D. E. Moser, Coordinated U–Pb geochronology, trace element, Ti-in-zircon thermometry and microstructural analysis of Apollo zircons. *Geochimica et Cosmochimica Acta* **202**, 264-284 (2017).
15. M. Grange, A. Nemchin, R. Pidgeon, The effect of 1.9 and 1.4 Ga impact events on 4.3 Ga zircon and phosphate from an Apollo 15 melt breccia. *Journal of Geophysical Research: Planets* **118**, 2180-2197 (2013).

16. P. H. Warren, G. J. Taylor, K. Keil, D. N. Shirley, J. T. Wasson, Petrology and chemistry of two "large" granite clasts from the Moon. *Earth and Planetary Science Letters* **64**, 175-185 (1983).
17. A. Nemchin, R. Pidgeon, M. Whitehouse, J. P. Vaughan, C. Meyer, SIMS U–Pb study of zircon from Apollo 14 and 17 breccias: implications for the evolution of lunar KREEP. *Geochimica et Cosmochimica Acta* **72**, 668-689 (2008).
18. A. Nemchin, M. Whitehouse, R. Pidgeon, C. Meyer, Oxygen isotopic signature of 4.4–3.9 Ga zircons as a monitor of differentiation processes on the Moon. *Geochimica et Cosmochimica Acta* **70**, 1864-1872 (2006).
19. M. Grange, R. Pidgeon, A. Nemchin, N. E. Timms, C. Meyer, Interpreting U–Pb data from primary and secondary features in lunar zircon. *Geochimica et Cosmochimica Acta* **101**, 112-132 (2013).
20. A. Nemchin *et al.*, Timing of crystallization of the lunar magma ocean constrained by the oldest zircon. *Nature geoscience* **2**, 133-136 (2009).
21. J. M. Mattinson, Zircon U–Pb chemical abrasion ("CA-TIMS") method: combined annealing and multi-step partial dissolution analysis for improved precision and accuracy of zircon ages. *Chemical Geology* **220**, 47-66 (2005).
22. M. Barboni *et al.*, Early formation of the Moon 4.51 billion years ago. *Science advances* **3**, e1602365 (2017).
23. A. Pourmand, N. Dauphas, Distribution coefficients of 60 elements on TODGA resin: application to Ca, Lu, Hf, U and Th isotope geochemistry. *Talanta* **81**, 741-753 (2010).
24. J. Zhang, N. Dauphas, A. M. Davis, A. Pourmand, A new method for MC-ICPMS measurement of titanium isotopic composition: Identification of correlated isotope anomalies in meteorites. *Journal of Analytical Atomic Spectrometry* **26**, 2197-2205 (2011).
25. C. Münker, S. Weyer, E. Scherer, K. Mezger, Separation of high field strength elements (Nb, Ta, Zr, Hf) and Lu from rock samples for MC-ICPMS measurements. *Geochemistry, Geophysics, Geosystems* **2** (2001).
26. T. Iizuka, T. Yamaguchi, Y. Hibiya, Y. Amelin, Meteorite zircon constraints on the bulk Lu–Hf isotope composition and early differentiation of the Earth. *Proceedings of the National Academy of Sciences* **112**, 5331-5336 (2015).
27. H. Gerstenberger, G. Haase, A highly effective emitter substance for mass spectrometric Pb isotope ratio determinations. *Chemical geology* **136**, 309-312 (1997).
28. D. Szymanowski, B. Schoene, U–Pb ID-TIMS geochronology using ATONA amplifiers. *Journal of Analytical Atomic Spectrometry* **35**, 1207-1216 (2020).
29. U. Söderlund, P. J. Patchett, J. D. Vervoort, C. E. Isachsen, The $^{176}$Lu decay constant determined by Lu–Hf and U–Pb isotope systematics of Precambrian mafic intrusions. *Earth and Planetary Science Letters* **219**, 311-324 (2004).
30. T. Hayakawa, T. Shizuma, T. Iizuka, Half-life of the nuclear cosmochronometer $^{176}$Lu measured with a windowless 4π solid angle scintillation detector. *Communications Physics* **6**, 299 (2023).
31. J. N. Connelly *et al.*, The absolute chronology and thermal processing of solids in the solar protoplanetary disk. *Science* **338**, 651-655 (2012).

32. A. Bouvier, J. D. Vervoort, P. J. Patchett, The Lu–Hf and Sm–Nd isotopic composition of CHUR: constraints from unequilibrated chondrites and implications for the bulk composition of terrestrial planets. *Earth and Planetary Science Letters* **273**, 48-57 (2008).
33. P. Sprung, T. Kleine, E. E. Scherer, Isotopic evidence for chondritic Lu/Hf and Sm/Nd of the Moon. *Earth and Planetary Science Letters* **380**, 77-87 (2013).
34. A. M. Gaffney, L. E. Borg, A young solidification age for the lunar magma ocean. *Geochimica et Cosmochimica Acta* **140**, 227-240 (2014).
35. T. Harrison *et al.*, Heterogeneous Hadean hafnium: evidence of continental crust at 4.4 to 4.5 Ga. *Science* **310**, 1947-1950 (2005).
36. M. Barboni *et al.*, High precision U-Pb zircon dating identifies a major magmatic event on the Moon at 4.338 Ga. *Science Advances* **10**, eadn9871 (2024).
37. L. E. Borg *et al.*, Isotopic evidence for a young lunar magma ocean. *Earth and Planetary Science Letters* **523**, 115706 (2019).
38. L. E. Borg, R. W. Carlson, The evolving chronology of Moon formation. *Annual Review of Earth and Planetary Sciences* **51**, 25-52 (2023).
39. N. Marks, L. Borg, I. Hutcheon, B. Jacobsen, R. Clayton, Samarium-neodymium chronology of an Allende calcium-aluminum-rich inclusion with implications for $^{146}$Sm isotopic evolution. *Earth Planet. Sci. Lett* **405**, 15-24 (2014).
40. L. E. Borg, A. M. Gaffney, C. K. Shearer, A review of lunar chronology revealing a preponderance of 4.34–4.37 Ga ages. *Meteoritics & Planetary Science* **50**, 715-732 (2015).
41. Y. Amelin, E. Rotenberg, Sm–Nd systematics of chondrites. *Earth and Planetary Science Letters* **223**, 267-282 (2004).
42. L. Fang *et al.*, Half-life and initial Solar System abundance of $^{146}$Sm determined from the oldest andesitic meteorite. *Proceedings of the National Academy of Sciences* **119**, e2120933119 (2022).
43. C. Burkhardt *et al.*, A nucleosynthetic origin for the Earth's anomalous $^{142}$Nd composition. *Nature* **537**, 394-398 (2016).
44. L. Nyquist *et al.*, $^{146}$Sm-$^{142}$Nd formation interval for the lunar mantle. *Geochimica et Cosmochimica Acta* **59**, 2817-2837 (1995).
45. M. Boyet, R. W. Carlson, A highly depleted moon or a non-magma ocean origin for the lunar crust? *Earth and Planetary Science Letters* **262**, 505-516 (2007).
46. A. D. Brandon *et al.*, Re-evaluating $^{142}$Nd/$^{144}$Nd in lunar mare basalts with implications for the early evolution and bulk Sm/Nd of the Moon. *Geochimica et Cosmochimica Acta* **73**, 6421-6445 (2009).
47. J. E. Dickinson Jr, P. Hess, Zircon saturation in lunar basalts and granites. *Earth and Planetary Science Letters* **57**, 336-344 (1982).
48. P. H. Warren, J. T. Wasson, The origin of KREEP. *Reviews of Geophysics* **17**, 73-88 (1979).
49. G. A. Gualda, M. S. Ghiorso, R. V. Lemons, T. L. Carley, Rhyolite-MELTS: a modified calibration of MELTS optimized for silica-rich, fluid-bearing magmatic systems. *Journal of Petrology* **53**, 875-890 (2012).
50. L. J. Crisp, A. J. Berry, A new model for zircon saturation in silicate melts. *Contributions to Mineralogy and Petrology* **177**, 71 (2022).